\documentclass[a4paper,12pt]{article}
\pdfoutput=1
 \usepackage{feynmp-auto,expdlist,multirow}
\usepackage{amsmath, amsfonts, amssymb,eufrak}
\usepackage[mathcal]{euscript}
\usepackage{graphicx}
\usepackage{enumerate}
\usepackage{hyperref}
\usepackage{latexsym}
\usepackage{hepnicenames}
\usepackage{enumerate}
\usepackage{soul}
\usepackage[normalem]{ulem}
 \usepackage{bbold}
\usepackage{units}
\usepackage{pdflscape}

\newcommand\LDL[1]{{\color{ForestGreen} [{\bf  LDL: #1}]}}

\newcommand{\g}{\mathfrak{g}}
\newcommand{\cgre}[1]{{\color{ForestGreen} #1}}

\newcommand{\A}{{\cal A}}
\newcommand{\B}{{\cal B}}

\newcommand{\C}{{\cal C}}
\newcommand{\G}{{\cal G}}

\newcommand{\W}{{\cal W}}
\newcommand{\Z}{{\cal Z}}
\newcommand{\X}{{\cal X}}
\renewcommand{\H}{{\cal  H}}
\renewcommand{\S}{{\cal S}}
\renewcommand{\P}{{\cal P}}
\newcommand{\N}{{\cal N}}
\newcommand{\M}{{\cal M}}
\newcommand{\tS}{{\tilde{\cal S}}}
\newcommand{\ts}{{\tilde{\s}}}
\newcommand{\ta}{{\tilde{\a}}}
\newcommand{\s}{\mathfrak{s}}
\renewcommand{\a}{\mathfrak{a}}
\newcommand{\x}{\mathfrak{x}}
\newcommand{\Dg}{\gamma_{\rm dark}}
\renewcommand{\N }{{\cal N}}

\usepackage{mathrsfs,graphicx,rotating,amsmath,amsfonts,mathtools,booktabs,amssymb,wasysym}
\usepackage{hyperref}\usepackage{slashed}
\usepackage[nosort]{cite}
\usepackage[table,xcdraw,dvipsnames]{xcolor}
\usepackage{bm}
\usepackage{graphicx}
\usepackage{multirow,multicol}
\hypersetup{colorlinks,bookmarksopen,bookmarksnumbered,
linkcolor=blus,pdfstartview=FitH,urlcolor=rossos,citecolor=verde}
\allowdisplaybreaks

\renewcommand{\[}{\left[}
\renewcommand{\]}{\right]}
\renewcommand{\(}{\left(}
\renewcommand{\)}{\right)}

\def\Lag{\mathscr{L}}

\newcommand{\mio}[1]{}

 \newcommand{\med}[1]{\langle #1\rangle}

\def\bpm{\begin{pmatrix}}
\def\epm{\end{pmatrix}}

\usepackage{mathrsfs}

 \newcommand{\fig}[1]{~\ref{fig:#1}}
\newcommand{\sfrac}[2]{#1/#2}

\renewcommand{\Re}{{\rm Re}\,}

\allowdisplaybreaks
\usepackage{multicol}
\usepackage{color}
\definecolor{rosso}{cmyk}{0,1,1,0.4}
\definecolor{rossos}{cmyk}{0,1,1,0.55}
\definecolor{rossoc}{cmyk}{0,1,1,0.2}
\definecolor{blu}{cmyk}{1,1,0,0.3}
\definecolor{blus}{cmyk}{1,1,0,0.6}
\definecolor{bluc}{cmyk}{1,1,0,0.1}
\definecolor{verde}{cmyk}{0.92,0,0.59,0.25}
\definecolor{verdec}{cmyk}{0.92,0,0.59,0.15}
\definecolor{verdes}{cmyk}{0.92,0,0.59,0.4}

\newcommand{\eq}[1]{~{\rm (\ref{eq:#1})}}

\newcommand{\GeV}{\,{\rm GeV}}
\newcommand{\TeV}{\,{\rm TeV}}

\newcommand{\Tr}{\,{\rm Tr}}
\newcommand{\diag}{\,{\rm diag}}

\def\circa#1{\,\raise.3ex\hbox{$#1$\kern-.75em\lower1ex\hbox{$\sim$}}\,}

\newcommand{\beq}{\begin{equation}}
\newcommand{\eeq}{\end{equation}}

\newcommand{\bea}{\begin{eqnarray}}
\newcommand{\eea}{\end{eqnarray}}
\newcommand{\be}{\begin{equation}}
\newcommand{\ee}{\end{equation}}
\font\tenrsfs=rsfs10 at 12pt
\font\sevenrsfs=rsfs7
\font\fiversfs=rsfs5
\newfam\rsfsfam
\textfont\rsfsfam=\tenrsfs
\scriptfont\rsfsfam=\sevenrsfs
\scriptscriptfont\rsfsfam=\fiversfs

\newcommand{\D}{{\cal D}}

\newcommand{\La}{\mathscr{L}}
\newcommand{\tr}{\mathrm{Tr}}
\newcommand{\I}{\mathbb{1}}

\newsavebox\MBox

\newcommand{\eV}{\,{\rm eV}}
\newcommand{\SU}{{\rm SU}}
\newcommand{\SO}{{\rm SO}}
\newcommand{\Sp}{{\rm Sp}}
\newcommand{\U}{{\rm U}}
\renewcommand{\L}{\mathscr{L}}

\def\circa#1{\,\raise.3ex\hbox{$#1$\kern-.75em\lower1ex\hbox{$\sim$}}\,}
\makeatletter

\font\ital=cmu10

\def\hhref#1{\href{http://arxiv.org/abs/#1}{arXiv:#1}}
\usepackage{xstring}
\newcommand{\hhrefq}[1]{\IfSubStr{#1}{:}{\href{http://inspirehep.net/search?ln=en&ln=en&p=#1&of=hb&action_search=Search&sf=&so=d&rm=&rg=25&sc=0}{InSpire:#1}}{\hhref{#1}}}

\def\art{\@ifnextchar[{\eart}{\oart}}
\def\eart[#1]#2#3#4#5#6{{\rm #2}, {\em #3 \bf #4} {\rm (#6) #5} ({\em #1})}
\def\article{\@ifnextchar[{\earticle}{\oarticle}}
\def\oarticle#1#2#3#4#5#6{{\rm #1}, {\ital ``#6''}, {\rm #2 #3 (#5) #4}}
\def\earticle[#1]#2#3#4#5#6#7{{\rm #2}, {\ital ``#7''}, {\rm #3 #4 (#6) #5}  [\hhrefq{#1}]}
\def\hepart[#1]#2{{\rm #2, \sl#1}}
\def\heparticle[#1]#2#3{#2, {\ital ``#3''} [\hhrefq{#1}]}
\newcommand{\doi}[1]{\href{http://dx.doi.org/#1}{[link]}}

\newcommand{\hhrefqq}[1]{\IfBeginWith{#1}{10.}{\href{https://doi.org/#1}{doi:#1}}{\hhrefq{#1}}}
\def\earticle[#1]#2#3#4#5#6#7{{\rm #2}, {\ital ``#7''}, {\rm #3 #4 (#6) #5}  [\hhrefqq{#1}]}

\renewenvironment{thebibliography}[1]
     {\begin{multicols}{2}[\section*{\refname}]%
      \@mkboth{\MakeUppercase\refname}{\MakeUppercase\refname}%
      \list{\@biblabel{\@arabic\c@enumiv}}%
           {\settowidth\labelwidth{\@biblabel{#1}}%
            \leftmargin\labelwidth
            \advance\leftmargin\labelsep
            \@openbib@code
            \usecounter{enumiv}%
            \let\p@enumiv\@empty
            \renewcommand\theenumiv{\@arabic\c@enumiv}}%
      \sloppy
      \clubpenalty4000
      \@clubpenalty \clubpenalty
      \widowpenalty4000%
      \sfcode`\.\@m}
     {\def\@noitemerr
       {\@latex@warning{Empty `thebibliography' environment}}%
      \endlist\end{multicols}}

\newcounter{alphaequation}[equation]
\def\thealphaequation{\theequation\hbox to
0.6em{\hfil\alph{alphaequation}\hfil}}
\def\eqnsystem#1{
\def\@eqnnum{{\rm (\thealphaequation)}}
\def\@@eqncr{\let\@tempa\relax \ifcase\@eqcnt \def\@tempa{& & &} \or
  \def\@tempa{& &}\or \def\@tempa{&}\fi\@tempa
  \if@eqnsw\@eqnnum\refstepcounter{alphaequation}\fi
\global\@eqnswtrue\global\@eqcnt=0\cr}
\refstepcounter{equation} \let\@currentlabel\theequation \def\@tempb{#1}
\ifx\@tempb\empty\else\label{#1}\fi
\refstepcounter{alphaequation}
\let\@currentlabel\thealphaequation
\global\@eqnswtrue\global\@eqcnt=0 \tabskip\@centering\let\\=\@eqncr
$$\halign to \displaywidth\bgroup \@eqnsel\hskip\@centering
$\displaystyle\tabskip\z@{##}$&\global\@eqcnt\@ne
\hskip2\arraycolsep\hfil${##}$\hfil& \global\@eqcnt\tw@\hskip2\arraycolsep
$\displaystyle\tabskip\z@{##}$\hfil
\tabskip\@centering&\llap{##}\tabskip\z@\cr}
\def\endeqnsystem{\@@eqncr\egroup$$\global\@ignoretrue} \makeatother

\definecolor{Gray}{gray}{0.95}

\def\bal#1\eal{\begin{align}#1\end{align}}

\begin{document}
\vspace{1.5cm}

\begin{center}
{\Large\LARGE\Huge \bf \color{rossos}
Scalar gauge dynamics\\[4mm] and Dark Matter}\\[1cm]
{\bf Dario Buttazzo$^{a}$, Luca Di Luzio$^{a,b,c}$, Parsa Ghorbani$^{a,b}$, Christian Gross$^{a,b}$,\\ Giacomo Landini$^{a,b}$,
Alessandro Strumia$^{b}$, Daniele Teresi$^{a,b}$, Jin-Wei Wang$^{b,d,e}$}\\[7mm]

{\it $^a$ INFN, Sezione di Pisa, Italy}\\[1mm]
{\it $^b$ Dipartimento di Fisica dell'Universit{\`a} di Pisa}\\[1mm]
{\it $^c$ DESY, Notkestra\ss e 85, D-22607 Hamburg, Germany}\\[1mm]
{\it $^d$ Key Laboratory of Particle Astrophysics, Institute of High Energy Physics,
	Chinese Academy of Sciences, Beijing, China}\\[1mm]
{\it $^e$ School of Physical Sciences, University of Chinese Academy of Sciences, Beijing, China}\\[1mm]

\vspace{0.5cm}

\begin{quote}\large
We consider theories with one gauge group (SU, SO or Sp)
and one scalar in a two-index representation.
The renormalizable action often has accidental symmetries
(such as global U(1) or unusual group parities)
that lead to one or more stable states, 
providing Dark Matter candidates.
We discuss the confined phase(s) of each theory and compute the two Higgs phases,
finding no generic dualities among them.
Discrete gauge symmetries can arise and
accidental symmetries can be broken, possibly giving pseudo-Goldstone Dark Matter.
Dark Matter candidates can have a complicated sub-structure characteristic of each group
and can be accompanied by extra dark radiation.
\end{quote}

\thispagestyle{empty}
\bigskip

\end{center}

\setcounter{footnote}{0}

\newpage

\tableofcontents


\section{Introduction}
We study the possibility that Dark Matter (DM) originates from elementary scalar/gauge dynamics.
We consider one elementary scalar $\S$
that fills a representation under a gauge group $\G$ with vectors $\mathcal{G}^A_{\mu}$
neutral under the SM gauge group.
We write the most generic renormalizable action, and study possible accidental symmetries
related to group theory that can lead to stable DM candidates.
Gauge theories predict non-trivial dynamics, leading to DM candidates with
non-minimal  cosmological history and specific features that depend on the gauge group.

The dark group $\G$ can become strongly interacting (`confined phase')
and/or get spontaneously broken by vacuum expectation values of $\S$
(`Higgsed phase' breaking $\G$ to a sub-group $\H$ that can confine at low energy).
A surprising equivalence between  the confined and Higgsed phases holds for
scalars $\S$ in the fundamental representation of the  groups
$\G=\{\SU(\N), \SO(\N), \Sp(\N), G_2\}$~\cite{Buttazzo:2019iwr}.
In these models $\S$ has a unique self-quartic, that
leads to a unique symmetry breaking pattern where
the only surviving scalar is a Higgs-like singlet.
We here extend the analysis considering a scalar in those representations such
that $\G$ is asymptotically free for any $\N$:
the symmetric, the anti-symmetric and the adjoint.\footnote{Furthermore, these  representations describe geometric configurations of
$\N$ self-intersecting $D$-branes~\cite{1711.04656}.}

%


The possible patterns of symmetry breaking of SU and SO groups
have been classified in~\cite{Li:1973mq} (see also~\cite{Wu:1981eb})
assuming renormalizable potentials.
We will extend these classical works by considering 
Coleman-Weinberg potentials, where running quartics lead
to dynamical symmetry breaking.
Furthermore, we consider SU, SO and Sp groups,
and symmetry breaking patterns that connect them.

In these theories
the scalar $\S$ has a mixed quartic coupling $\lambda_{H\S}$ with the Higgs and
two different quartic self-couplings.
This leads to two different patterns of symmetry breaking and to
extra scalars in the broken phases.\footnote{The possibility that DM is a composite state made of elementary fermions
in QCD-like theories was explored in~\cite{1503.08749,1707.05380}.
As elementary scalars interact with the Higgs through a scalar quartic,
while elementary fermions can have Yukawa couplings, the phenomenology is different.
In particular the elementary scalar can be neutral under the SM gauge group,
such that collider experiments can probe the
dark strong sector 
through precision Higgs measurements.}


In section~\ref{General} we discuss some common generic features:
the breaking patterns; the possible accidental symmetries intrinsic of
the SU, SO, Sp gauge groups; the possible equivalence between the Higgs and confined phases.
Furthermore, dynamical symmetry breaking 
through scalars in adjoints  leads to unbroken U(1) factors: 
in section~\ref{Dg} we summarize the phenomenology
of DM charged under a dark U(1).
Dark monopoles become possible DM candidates, as summarized in section~\ref{darkMono}.

\medskip

We next discuss the concrete models:
a symmetric of $\SU(\N)$ in section~\ref{SUNsym},
an anti-symmetric of $\SU(\N)$ in section~\ref{SUNantisym},
an adjoint of $\SU(\N)$ in section~\ref{SUNadj},
a trace-less symmetric of $\SO(\N)$ in section~\ref{SONsym},
an anti-symmetric adjoint of $\SO(\N)$ in section~\ref{SONantisym},
a symmetric adjoint of $\Sp(\N)$ in section~\ref{SpNsym},
a trace-less anti-symmetric of $\Sp(\N)$ in section~\ref{SpNantisym}.
These cases need to be discussed separately as each one has its specific features.
In each case we discuss the accidental symmetries, the renormalization group equations (RGE),
the confined phase, the Higgs phases
(in particular we compute assuming dynamical symmetry breaking \`a la Coleman-Weinberg \cite{Coleman:1973jx}),
and the DM candidates in each phase.
We summarize our results and the generic lessons in the
conclusions, section~\ref{concl}.

\section{Accidental symmetries}\label{General}

\begin{table}[t]
\begin{center}\footnotesize
\begin{tabular}{l|c|c|c}
\rowcolor[cmyk]{0,0,0.2,0}&\multicolumn{3}{c}{Unbroken group and its dimension}\\ \hline
\rowcolor[cmyk]{0.1,0,0.1,0}Representation & $\SU(\N),~\N^2-1$ & $\SO(\N),~\N(\N-1)/2$ & $\Sp(\N),~\N(\N+1)/2$ \\
\hline \hline
fundamental & $\SU(\N-1)$ & $\SO(\N-1)$ & $\Sp(\N-2)$ \\ \hline
symmetric &  $ \SU(\N-k)\otimes\SO(k)$ & $\SO(\N-k) \otimes \SO(k)$ & $\Sp(\N-2k)\otimes\SU(k)\otimes{\rm U}(1)$ \\ \hline
anti-symmetric & $ \SU(\N-2k)\otimes \Sp(2k)$ & 
$\SO(\N-2k)\otimes\SU(k) \otimes {\rm U}(1)$ & $\Sp(\N-2k)\otimes\Sp(2k)$\\ \hline
adjoint & $\SU(\N-k) \otimes \SU(k) \otimes {\rm U}(1)$ &
\hbox{see anti-symmetric} & see symmetric
 \\
\hline
\end{tabular}
\end{center}
\caption{\em\label{tab:breakpatterns} 
Breaking patterns with one irreducible scalar representation
of small size, such that the gauge group confines for any $\N$.
The symmetric of $\SO(\N)$ is trace-less,
with a similar condition on the anti-symmetric of $\Sp(\N)$.
The values of $k$ at the absolute minima of renormalizable potentials~\cite{1711.04656} and of Coleman-Weinberg potentials
are either minimal  ($k=1$) or maximal ($k=\N$ or $\N-1$).
We recall that  $\SU(2)=\SO(3)=\Sp(2)$,
$\SO(5)=\Sp(4)$, $\SU(4)=\SO(6)$, $\SO(4)=\SU(2)^2$.
}
\end{table}

\subsection{Gauge symmetry breaking patterns}
Table~\ref{tab:breakpatterns} lists the breaking patterns
produced by one scalar 
in one 1-index or 2-index representation of $\SU$, $\SO$, $\Sp$ groups, 
such that the gauge beta function is asymptotically free for any $\N$. 
That would not be the case, for instance, for a 3-index representation
(whose 
contribution to the gauge beta function scales as $\N^3$)
or for an SO spinor (scaling as $\sim 2^{\N/2}$).
The $\SU$ and $\SO$ breaking patterns were 
classified in~\cite{Li:1973mq,Wu:1981eb,Jetzer:1983ij} that assumed
a quartic renormalizable potential. 
The extension to $\Sp$ groups was recently considered in~\cite{1711.04656}.
We also consider scale-invariant potentials, where running quartics
lead to dynamical symmetry breaking  \`a la Coleman-Weinberg. 

In section~\ref{acc} we discuss the possible group-theoretical accidental
global symmetries that can lead to a stable DM candidate. According to unknown qualitative aspects of strong dynamics, some of these symmetries could be, in principle, broken by the formation of scalars condensates, as we discuss in section~\ref{VWlike}.

\subsection{Accidental symmetries}\label{acc}
First, we need to know which representations are real and which are complex.
With this information, we can next write actions and identify their accidental symmetries.

\subsubsection*{Reality conditions}
The generators $T$ acting on real and pseudo-real representations  satisfy 
$T^* = - V^{-1} T V$, where $V$  is a symmetric matrix for real representations (e.g.~$V=\I$ 
for the $\N$ of
$\SO(\N)$) and anti-symmetric for pseudo-real representations
(e.g.~$V=\gamma_\N \equiv \diag(\epsilon,\ldots,\epsilon)$ for the $\N$ of $\Sp(\N)$, 
with $\epsilon = i \sigma_2$ in terms of Pauli matrices). 
Then, a scalar in the fundamental representation $\S_I$  transforms in the same way as $(V \S^*)_I$. A non-trivial reality condition on $\S_I$, i.e.~$\S = V \S^*$, can be then imposed only if $VV^*=\I$,
 i.e.~for real groups but not for pseudo-real (or complex) ones. Instead, a reality condition can be imposed on a 2-index representation $\S_{IJ}$ (adjoint, symmetric, anti-symmetric or bi-fundamental), 
i.e.~$\S^* = - V^{-1} \S V$,
 for both real and pseudo-real groups, since $VV^* = \pm \I$. 
To summarize, the elements of the fundamental of $\SU(\N)$ and $\Sp(\N)$ and the symmetric and the anti-symmetric of $\SU(\N)$ need to be complex numbers. 

\subsubsection*{Global U(1)}
The renormalizable action of a complex scalar
can be invariant under an accidental global $\U(1)$ symmetry that acts as an overall rephasing of $\S$. 
The lightest state charged under this U(1) is a DM candidate
(extra co-stable states are possible, depending on the spectrum).
When a SU (Sp) gauge interaction confines, the global U(1) results into stable baryons (mesons).

\subsubsection*{Local U(1) and $\mathbb{Z}_2$}
Some symmetry breaking patters $\G\to \H$ leave
unbroken a gauged $\U(1)$ or $\mathbb{Z}_2$ sub-group of $\G$
that can imply DM stability.
In the simplest case, a $\mathbb{Z}_2$ arises when $\G={\rm U}(1)$ is broken
by a scalar with charge 2.
We will consider non-abelian groups with scalars in two-index representations
finding more examples of such $\mathbb{Z}_2$ (see e.g.~section~\ref{Sp1Sp2}).
We loosely include such $\mathbb{Z}_2$ among the accidental symmetries,
despite that they are gauge discrete symmetries.

\subsubsection*{Group parities $\P$}
The elements of the adjoint of $\SU(\N)$ and all the 2-index representations of $\SO(\N)$ and $\Sp(\N)$ (including bi-fundamentals of two $\Sp$ groups) can be taken to be real, so that there is no global $\U(1)$.  
These theories can be accidentally invariant under an
accidental ``group parity'' $\mathbb{Z}_2$ discrete symmetry.
Unlike usual $\mathbb{Z}_2$ symmetries, 
this symmetry acts on components of multiplets rather than on multiplets.
It is analogous to the usual space-parity, except that  it acts in group space
rather than in space.

Group parity can be an accidental symmetry for $\SU(\N)$ and $\SO(\N)$ groups because
it is broken only by terms involving the Levi-Civita $\epsilon$ anti-symmetric tensor with $\N$ indices:
for $\N$ large enough, $\epsilon$ does not appear in the renormalizable action.
On the other hand, for $\Sp(\N)$ groups, possibile group parities are broken by terms involving the
$\gamma_\N$ tensor (analogous of the $\delta$ tensor). 
As it has just two indices, the action contains terms odd under parity.

For $\SU(\N)$, the theory is accidentally invariant under a reflection,
which we dub U-parity, of any of the $\N$ equivalent directions in group space.
U-parity is obtained by flipping the sign of any color, for example the 1st one.
This flips the signs of those generators with an $1I$ entry,
preserving the $\SU(\N)$ Lie algebra, such that U parity acts on
components of vectors in the adjoint and of other multiplets
as
\be\label{eq:Uparity}
\G_I^J \stackrel{\P_\text{U}}\to (-1)^{\delta_{1I} + \delta_{1J}} \G_I^J,
\qquad 
{\cal S}_{I} \stackrel{\P_{\rm U}}\to (-1)^{\delta_{1I}}  {\cal S}_{I} ,\qquad
{\cal S}_{IJ} \stackrel{\P_{\rm U}}\to (-1)^{\delta_{1I}+\delta_{1J}}  {\cal S}_{IJ} 
\ee 
having written the $\SU$ vectors $\G_{J}^I =\G^A (T^A)_J^I$.

For $\SO(\N)$, the theory is accidentally invariant under a reflection under any of the $\N$ equivalent directions in group space: the resulting accidental $\mathbb{Z}_2$ symmetry is O-parity \cite{Witten:1983tx,Buttazzo:2019iwr}.
Representing O-parity by flipping the direction 1, its action is
analogous to eq.\eq{Uparity}, dropping the distinction between
indices in the fundamental and anti-fundamental:
\beq
\G_{IJ} \stackrel{\P_{\rm O}}{\to} (-1)^{\delta_{1I}+\delta_{1J}} \G_{IJ},\qquad
{\cal S}_{I} \stackrel{\P_{\rm O}}\to (-1)^{\delta_{1I}}  {\cal S}_{I} ,\qquad
\qquad  {\cal S}_{IJ} \stackrel{\P_{\rm O}}\to (-1)^{\delta_{1I}+\delta_{1J}}  {\cal S}_{IJ} .
\eeq
When a SU or SO gauge group confines, 
dark baryons built contracting constituents 
with one $\epsilon$ tensor
are odd under group parity. 
The lightest odd state can be a stable DM candidate.

\subsubsection*{Group charge conjugations $\C$}
Theories with  $\SU(\N)$ gauge groups can be invariant under
a charge conjugation symmetry.
{As a result, $\SU(\N)$ vectors with purely real generators are odd, while
vectors with purely imaginary generators are even}.
Theories with  $\SO(\N)$ or $\Sp(\N)$ gauge groups (with $\N$ even) can be invariant under
a charge conjugation symmetry that extends the one of their $\U(\N/2)$ subgroup. For $\SO$ this has been studied in \cite{Buttazzo:2019iwr}: the only invariant tensor that can give rise to odd states is again $\epsilon$, so that it does not give rise to new stable states. 

A similar situation is found for $\Sp(\N)$, as we now discuss (although the discussion is more complicated and will have no practical relevance).
The symmetry, that  we dub $\C_{\Sp}$, 
acts on the 2-index representations considered in this paper as
\be \label{eq:transfCSp}
\G_{I}^{J} \to (-1)^{I+J} \G_{I}^{J}, \qquad \S_{I}^{J} \to (-1)^{I+J} \S_{I}^{J}.
\ee
Let us now show that this is a symmetry of the action. 
That is trivially true for bilinears and quartics;
cubics $\S^3$, when present, respect the symmetry\footnote{This crucially depends on the overall sign of the transformation of $\S$ chosen in~\eqref{eq:transfCSp}. Notice that, as a consequence, one has $\hat\S_{IJ} \equiv ( \S \gamma_\N)_{IJ} \to - (-1)^{I+J} \hat\S_{IJ}$.} because all indices are contracted, so they appear an even number of times. Then, one only needs to check the compatibility with the Lie algebra of $\Sp(\N)$. Let us denote by $T_{\rm sym} = \{T_{\rm real}, \I_{\N/2}/\sqrt{\N}\}$ the symmetric generators of $\U(\N/2)$, and by 
$T_{\rm asym} = T_{\rm imag}$ the anti-symmetric ones. 
Among the $\Sp(\N)$ generators,\footnote{We remind the reader that the $\Sp(\N)$ generators can be written as 
\beq 
\frac{T_{\rm asym} \otimes \I_2}{\sqrt{2}}, \qquad \frac{T_{\rm sym} \otimes \sigma_k}{\sqrt{2}}
\eeq
where $k=1,2,3$. In the notation of Appendix A of \cite{Buttazzo:2019iwr}, 
$T_{\rm asym} = T_{\alpha \beta}^{(2)}$ and 
$T_{\rm sym} = \{ T_{\alpha \beta}^{(1)}, \frac{1}{\sqrt{2}} T_{\alpha \alpha}^{(1)} \}$, 
with $\alpha \neq \beta$. 
} 
those of the form 
\be\label{eq:evengen}
\frac{T_{\rm asym} \otimes \I_2}{\sqrt{2}}, \qquad \frac{T_{\rm sym} \otimes \sigma_3}{\sqrt{2}}
\ee
are even under $\C_{\Sp}$ (because the $2\times 2$ blocks are diagonal), whereas 
\be\label{eq:oddgen}
\frac{T_{\rm sym} \otimes \sigma_1}{\sqrt{2}}, \qquad \frac{T_{\rm sym} \otimes \sigma_2}{\sqrt{2}}
\ee
are odd (because the $2\times 2$ blocks are off-diagonal). The Lie algebra is compatible with this symmetry: the product of two blocks in the set $\{\I_2, \sigma_3\}$ stays in the set, so that $[T_{\rm even}, T_{\rm even}]$ is even; the product of two blocks in the set $\{\sigma_1, \sigma_2\}$ is in  $\{\I_2, \sigma_3\}$, so that $[T_{\rm odd}, T_{\rm odd}]$ is even; the product of two blocks belonging to the two different sets is in $\{\sigma_1, \sigma_2\}$, so that $[T_{\rm even}, T_{\rm odd}]$ is odd.

As anticipated above, this symmetry is nothing but charge conjugation $\cal{C}$ for the $\U(\N/2)$ subgroup of $\Sp(\N)$: since $\U(\N/2) = \SO(\N) \cap \Sp(\N)$, the $\U(\N/2)$ sub-algebra is given by the anti-symmetric generators, i.e.\ the first set in \eqref{eq:evengen} and the second set in $\eqref{eq:oddgen}$; 
the imaginary (real) generators are even (odd) under charge conjugation of $\U(\N/2)$, and the same happens under the symmetry of $\Sp(\N)$.

The only invariant tensor that gives rise to odd states is $\gamma_\N$. 
If $\S$ is complex, $\C_{\Sp}$ implies stability of mesons which are
already stable because of accidental U(1).
If $\S$ is real and a two-index representation, $\C_{\Sp}$ does not imply new stable bound states. $\C_{\Sp}$ could give rise to new stable states for real 1-index $\S$, such as the bi-fundamental arising in the Higgs phases in section \ref{SpNantisym}. However, the $\C_{\Sp}$ of the subgroups appearing there are broken in the full theory. To summarize, $\C_{\Sp}$ never gives rise to new stable states in the cases discussed in this work, analogously to $\C_{\SO}$. 



\subsection{Does scalar gauge dynamics break accidental symmetries?}\label{VWlike}
Ref.~\cite{Buttazzo:2019iwr} discussed the possible theories 
of scalars with gauge interactions such that
the confined phase is self-dual to the Higgs phase. 
In all those theories the scalar $\S$ is in the fundamental representation of the gauge group $\G$
such that the breaking  $\G\to\H$ is univocally fixed by group theory.
The conjectured duality is based on the matching between 
the spectra and accidental symmetries in the two phases:
each particle in the Higgs phase, after condensation of the subgroup $\H$, is associated to a corresponding operator {\em invariant} under $\G$, 
so that one expects that the Higgs and confined phases are smoothly connected in the 
strong coupling limit $\g \sim 4\pi$.

We here study theories with scalar content such that
$\G$ can be perturbatively broken to two different sub-groups, $\H_1$ or $\H_2$.

At first sight, an apparently related result is the Fr\"ohlich-Morchio-Strocchi (FMS) 
theorem~\cite{Frohlich:1981yi}.
Seeking a manifestly gauge covariant description of the Higgs phase, 
FMS prove that for each state in the Higgs phase 
there is a corresponding operator in the unbroken theory {\em covariant} under $\G$.\footnote{However, we find that in some situations the gauge-covariant description of the Higgs phase cannot be performed by means of finite polynomials of fields. Take for instance an $\SU(\N)$ theory with a scalar $\S$ in the symmetric that breaks it to $\SU(\N-1)$ and a fermion $\psi$ in the fundamental. In the Higgs phase, the fermion $\psi$ splits into a singlet $\psi_0$ and a fundamental of $\SU(\N-1)$. According to the FMS theorem, the fermion singlet $\psi_0$ can be described by a singlet of $\SU(\N)$. However, for even $\N$, no finite polynomial field operators  are fermionic and singlet under $\SU(\N)$.}
In some theories the corresponding operators are invariant singlets,
pointing to a relation between the confined and Higgs phases.
But in general, the corresponding operators are only covariant:
their existence does not imply a duality between the Higgs and confined phases.
The formation of vacuum expectation values and/or condensates, 
even when described by means of singlets of the original group $\G$
(without referring to the sub-group $\H$) is a dynamical phenomenon
not controlled by the group-theoretical FMS theorem.

\smallskip

Therefore, whether the Higgs and confined phases are dual is a non-trivial question. 
In particular, dynamics could form complicated condensates that break the accidental symmetries of the theory
and invalidate gauge/Higgs dualities.
Let us consider, for example, a confined phase where a baryon $\B$ is stable because of an accidental $\U(1)$. 
If the condensate $\langle \B \rangle$ forms in some region of the parameter space,
it would violate the accidental symmetry, destabilising the DM candidate stable thanks  to baryon number,
and giving instead rise to a massless Goldstone boson. 
An analogous situation could arise in the Higgs phase, when non-abelian factors of the 
unbroken sub-group $\H$ become strongly interacting and generate condensates.
If U(1)-breaking condensates form only in one of the two phases (purely condensed, or Higgsed),
their duality would be lost.

For a vector-like fermion-gauge theory, a Vafa-Witten (VW) theorem~\cite{Vafa:1983tf} guarantees that massless Goldstone bosons of vector-like symmetries are not present in the spectrum, 
and hence that baryon $\U(1)$-violating condensates do not form. 
The VW theorem does not hold in the presence
of a topological term $\theta \G\tilde \G$
(as it makes the  Euclidean path integral not real)
nor in the presence of non-gauge interactions, such as scalar self-interactions. A trivial counter-example is usual spontaneous symmetry breaking in the limit of vanishing gauge coupling. 
However, the dynamics behind the VW theorem suggests that strong
gauge interactions tend not to break global vector-like symmetries 
(somehow analogously to how gauge interactions tend to maximise the unbroken symmetries in
composite Higgs models). 

While scalars have extra quartic interactions that can behave differently,
we can heuristically expect that strong gauge interactions play the dominant role,
provided that the numerical value of scalar quartics at confinement is not too large.
One-loop RGE running tends to make quartics large and negative at lower energy,
but in the Higgs phases scalars avoid large quartics, being heavier than the scale
where  gauge couplings run non-perturbative.
If quartics do not alter significantly gauge dynamics, it seems plausible to assume 
that U(1)-violating condensates do not form at the condensation of the residual gauge group $\H$.
The situation in the confined phase is more uncertain, 
because quartics of massless scalars typically run to large values when approaching confinement.

Furthermore, the VW theorem relies on showing the absence of Goldstone modes in the spectrum, so that
it does not apply to discrete global symmetries such as 
those found in the previous section~\ref{acc}.
Discrete symmetries might be broken even by pure gauge dynamics. 
More in general, several different confined phases could be present, 
corresponding to the formation of different sets of condensates. 

For most of the discussion below we will assume that accidental symmetries (discrete or  continuous)
are not broken by the formation of condensates.
The reader should keep in mind that this could be true only in part of the parameter space.

\section{Models with unbroken dark U(1)}\label{Dg}
Some of the theories considered in the following leave an unbroken dark
gauged U(1) under which (part of) DM is charged.  
Such physics is constrained by
bounds on dark radiation (as summarised in section~\ref{darkRadiation}),
by dark Coulombian scatterings (section~\ref{darkCoulomb}),
and by dark matter/dark radiation interactions (section~\ref{gammaDark}),
and still allowed in regions of the parameter space where DM is heavy enough.
See also \cite{Balkin:2018tma}.

Furthermore, whenever a dark $\U(1)$ factor remains in the unbroken gauge group, 
dark monopoles are stable and contribute to the 
DM relic density (see e.g.~\cite{Murayama:2009nj,Baek:2013dwa,Khoze:2014woa};
see also \cite{Bryan:1993hz} for other possible topological states). 
This will be discussed in section~\ref{darkMono}.

\subsection{Dark radiation}\label{darkRadiation}
Cosmology provides information on the amount of energy density in relativistic species,
usually reported as an effective number of extra neutrino species. The Planck satellite results imply~\cite{1807.06209}
\beq\label{eq:DRbound}
\Delta N_{\rm eff} \leq 0.30 \qquad \hbox{at 95 \% C.L.}
\eeq
at the CMB temperature.
A dark photon with energy density $\rho_{\rm dark}$ contributes as
\beq  \Delta N_{\rm eff} = \frac{8}{7} \bigg(\frac{11}{4}\bigg)^{4/3} \frac{\rho_{\rm dark}}{\rho_\gamma}.
\eeq
We will consider models (specified later) where a dark scalar $\S$ 
interacts with the SM as $\lambda_{H\S} |H|^2|\S|^2$.
At a scale $w$ the dark scalar breaks a dark gauge group $\G$ to a U(1) times a non-abelian subgroup 
that confines at a scale $\Lambda$.

If the dark photon $\Dg$ 
decouples from the SM at a temperature $T_{\rm dec} \circa{<}\Lambda$ low enough that $\Dg$ is the only
dark-sector particle, it contributes as $g_{*s}^{\rm dark}(T_{\rm dec})=2$ such that
\beq\label{eq:deltaN}
\Delta N_{\rm eff}= \frac{8}{7} \(\frac{43 g_{*s}^{\rm dark}(T_{\rm dec})}{8 \, g_{*s}^{\rm SM}(T_{\rm dec})} \)^{\sfrac{4}{3}} 
= 0.058  \(\frac{100}{ g_{*s}^{\rm SM}(T_{\rm dec})} \)^{\sfrac{4}{3}} 
\eeq
which is allowed for $T_{\rm dec}$ above the QCD or electroweak scale.
We estimate $T_{\rm dec}$ as follows.
The thermal interaction rate of dark photons is $\Gamma \approx \lambda_{HS}^2 \sfrac{T^5}{M_s^4}$
where $M_s$ is the mass of the dark Higgs.
Imposing $\Gamma \sim H \sim T^2/M_{\rm Pl}$ gives
$ T_{\rm dec} \sim \sfrac{M_s^{4/3}}{M_{\rm Pl}^{1/3} \lambda_{HS}^{2/3}}$.
Our initial assumption $T_{\rm dec} \circa{<} \Lambda$ is satisfied for
\beq 
\frac{\Lambda}{M_s} \gtrsim \( \frac{M_s}{\lambda_{HS}^2 M_{\rm Pl}}\)^{\sfrac{1}{3}} \;.
\eeq
If instead dark photons decouple at $T_{\rm dec}\gg \Lambda$
when the dark sector contains dark gluons $\A$ as extra degrees of freedom,
the bound of eq.\ \eqref{eq:DRbound} is violated (for any SU(2) or bigger confining group)
if the extra energy ends up reheating dark photons
rather than the SM.
This conclusion can only be avoided if the dark glue-balls mostly decay out of equilibrium
into SM particles rather than into dark photons.




Let us estimate the decay widths of dark glue-balls.
Dark glue-balls can decay into the dark photons only through dimension-8 operators
\be\label{eq:dim8}
C_1(\A_{\mu\nu}^a)^2(\Dg^{\rho\sigma})^2 +C_2 (\A_{\mu\nu}^a\Dg^{\mu\nu})^2.
\ee
with $C_i \sim 1/(4\pi w^2)^2$.
The glue-ball decay widths are  estimated as (omitting factors of $\g\sim 1$)
\beq
\Gamma_{\rm SM} \sim   \frac{\lambda_{HS}^2 \Lambda^5}{(4\pi)^4 M_s^4} \, , \qquad \Gamma_{\rm dark} \sim  \frac{\Lambda^9}{(4\pi)^4w^8}
\eeq
where $w$ is the vacuum expectation value that breaks the dark gauge group $\G$.
Assuming that the dark-sector contribution to the Higgs squared mass is not unnaturally large,
$\lambda_{HS} \circa{<} v^2/w^2$, 
we estimate that glue-balls dominantly decay into SM particles if $\Lambda \circa{<} v$.

\begin{figure}
$$\includegraphics[width=0.45\textwidth]{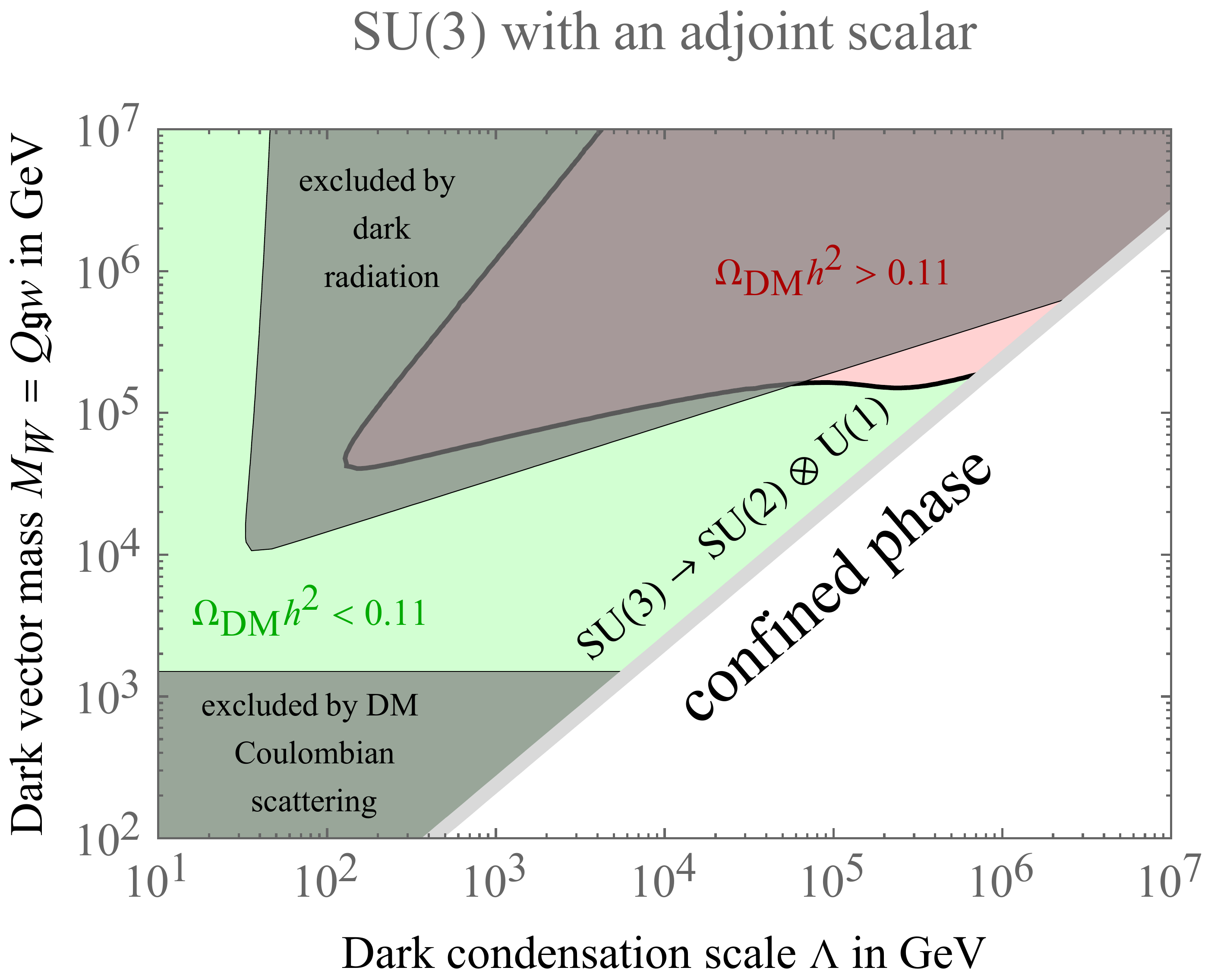}$$
\caption{\em \label{fig:Dgplot} We consider a typical model with DM charged under dark photons:
one Higgs phase of gauged  $\SU(3)$ with a scalar in the adjoint.
The cosmological DM abundance is reproduced thermally around the boundary between
the green (DM under-density) and red (DM over-density) regions.
The regions in grey are excluded by the resulting dark radiation and Coloumbian scatterings
among DM particles.
The confined phase of this model (in white) features no DM candidate.
}
\end{figure}

\subsection{Coulombian scattering among DM particles}\label{darkCoulomb}
The massless dark photon $\Dg$ mediates Coulombian scattering among two DM particles.
We consider its phenomenology and bounds.
The transfer cross section (cross section weighted by fractional longitudinal momentum transfer)
is given by~\cite{0911.0422}
\beq \sigma_{\rm tran} = \frac{\g^4}{\pi M^2 v^4}\ell\eeq
where $\ell \sim$ tens is a logarithmic IR enhancement cut-offed by the small plasma mass
acquired by massless vectors in a cosmological DM background.
Various observations (the bullet cluster, black hole accretion, tri-axiality of galaxies, etc) 
demand
$\sigma_{\rm tran}/M \circa{<} {\cal O}({\rm cm^2/g}) $ at $v \sim (30-1000)\,{\rm km/s}$~\cite{1610.04611,1705.02358,1908.00207},
implying 
\beq \label{eq:Dgammabound}
M \circa{>} (0.3-3)\TeV\times \g^{4/3}.\eeq
When such bound is nearly saturated, self-interacting
DM  can marginally improve the potential 
core-vs-cusp and too-big-to-fail problems.

\subsection{Interactions of DM with the  dark photon}\label{gammaDark}
Cosmological data are reproduced assuming that DM freely streams
during structure formation at $T\circa{<} T_{\rm eq}\approx 0.74\eV$.
However, interactions in the dark sector can lead to a different fluid similar to the baryon/photon system.
This effect is controlled by the dark Thomson cross section
between DM and dark photons, $\sigma_T =\g^4/6\pi M^2$.
Dark photons decouple very early from DM, when $n_{\rm DM} \sigma_T \circa{<} H$.
DM decouples from dark photons when $n_{\Dg} \sigma_T (T/M) \circa{<} H$
(where the $T/M$ factor accounts for the needed energy transfer), at
\beq T_{\rm fs} \sim\frac{M^{3/2}}{\g^2 M^{1/2}_{\rm Pl} } \sim 
\frac{\GeV}{\g^2} \bigg(\frac{M}{100\TeV}\bigg)^{3/2}.\eeq
Imposing $T_{\rm fs} > T_{\rm eq}$ gives 
\beq M \circa{>} \g^{4/3} M_{\rm Pl}^{1/3} T_{\rm eq}^{2/3} \sim \g^{4/3}\GeV\eeq
which is weaker than eq.\eq{Dgammabound}.\footnote{Our result differs
even parametrically from the previous result in~\cite{0810.5126}.}


\subsection{Dark monopoles}\label{darkMono}
When the dark gauge group is broken to a subgroup that contains an unbroken U(1) factor,
dark magnetic monopoles exist with
magnetic charge $\g_{\rm mag} = 4 \pi / \g  $ and mass $M_{\rm mag} \sim M_{\W} / \alpha_{\rm DC}$.

Their abundance is negligible (one per Hubble volume at $T \sim M_\W$)
if symmetry breaking occurs through a first order phase transition 
(as in the Coleman-Weinberg case).
The monopole abundance can instead be significant 
\beq 
\Omega_{\rm mag} h^2 \approx 1.5 \times 10^{9}  \frac{M_{\rm mag}}{ \text{TeV}} 
\( \frac{30 \ T_{\rm cr}}{M_{\rm Pl}} \)^{\frac{3\nu}{1+\nu}} 
\xrightarrow[]{\nu = 1/2} 0.12 \frac{T_{\rm cr}}{3.2 \times 10^7 \ \text{GeV}} 
\eeq
if a second order phase transition takes place with critical temperature $T_{\rm cr}\sim M_\W$~\cite{Murayama:2009nj,Baek:2013dwa,Khoze:2014woa}.
If the critical exponent has the `classical' value $\nu=1/2$ and the gauge group is
strongly coupled, $\g\sim 4\pi$ at  $M\sim 100\TeV$, 
both $\W$ and dark magnetic monopoles have abundances comparable to the DM cosmological abundances.
Monopoles become unstable
if the U(1) is spontaneously broken at low-energy.

\section{A symmetric of SU($\N $)}\label{SUNsym}
We now consider a scalar $\S_{IJ}$ in the symmetric complex representation of SU($\N $).
The dimension-less Lagrangian for generic $\N \neq 2,4$ accidentally conserves
a U(1) dark baryon number and is
\be\label{eq:LagSym}
\La=\La_{\rm SM}-\frac{1}{4} \mathcal{G}^A_{\mu\nu}\mathcal{G}^{A\,\mu\nu}+
\Tr (\D_\mu\S)(\D^\mu\S)^\dag -V_\S\ee
with
\be \label{eq:VS}
V_\S= M_\S^2 \Tr(\S^\dagger\S)+\lambda_\S (\!\Tr \S\S^\dag)^2+
\lambda'_\S \Tr (\S\S^\dag \S\S^\dag)-
 \lambda_{H\S} |H|^2 \Tr \S\S^\dag .
\ee 
The cases $\N = 2,3,4$ are special because
$\det\S$ (invariant for any $\N $) gives an extra renormalizable term
that breaks the U(1) global symmetry.
For $\N =2$ it is a mass term and
a reality condition can be imposed, reducing the 
components to 3, such that only a quartic exists and no U(1) symmetry arises.
Indeed the symmetric of SU(2) is the adjoint (or fundamental of SO(3)).
For $\N=3$ the theory admits the extra complex cubic $\det\S$.
For $\N=4$ the theory admits the extra complex quartic 
\beq V_{\rm extra}= \lambda''_\S \det\S = \frac{\lambda''_\S }{4!}
\epsilon^{IJKL} \epsilon^{I'J'K'L'} \S_{II'} \S_{JJ'}\S_{KK'}\S_{LL'}.\eeq
It breaks dark baryon number, and it is thereby not generated by RGE.
Ignoring it, the RGE for $\N \ge 3$ are 
{\small
\begin{eqnsystem}{sys:RGEsym}
(4\pi)^2 \frac{d\g}{d\ln\mu} & = & -\frac{21\N -2}{6} \g^3\\
(4\pi)^2 \frac{d\lambda_\S}{d\ln\mu} &=&\g^4\left(9+\frac{24}{\N ^2}\right)-12 \g^2 \lambda _\S
\frac{\N ^2+\N -2}{\N }+\\ \nonumber
&&+6 \lambda^{\prime 2} _S+8 
   (\N +1) \lambda _\S \lambda' _\S+2 \left(8 +\N +\N ^2\right) \lambda _\S^2\\
   (4\pi)^2 \frac{d\lambda'_\S}{d\ln\mu} &=&
   3\g^4 \left(\N -\frac{16}{\N }+4\right)-12 \g^2  \lambda' _\S\frac{ \N ^2+\N -2}{\N }+2  \lambda^{\prime 2} _\S (2 \N +5)+24 \lambda_\S \lambda'_\S.
\end{eqnsystem}}

\subsection{A symmetric of SU($\N $): confined phase}\label{sec:SUNsymconf}
We define dark baryon number such that $\S$ has charge $2/\N $.
In the $\SU(\N )$-condensed phase the following hadrons charged under baryon number form:
\begin{itemize}
\item for $\N =2n$ even the baryon 
\beq
\B=\epsilon^{i_1 j_1\cdots i_{n} j_{n}} (\G_{\mu_1\nu_1})^{k_1}_{i_1} \S_{k_1j_1}\cdots 
(\G_{\mu_{n}\nu_{n}})^{k_{n}}_{i_{n}} \S_{k_{n}j_{n}}\eeq
For SU(2) = SO(3) the symmetric $\S_{IJ}$ equals  the fundamental of $\SO(3)$, and the baryon
reduces to the odd-ball of~\cite{Buttazzo:2019iwr}:
$\epsilon^{IJ} (\G_{\mu\nu})^K_I \S_{KJ} = \S^A\G^A_{\mu\nu}  = \epsilon^{ABC} \S^A \G^{BC}_{\mu\nu}$.

\item for any $\N $ the di-baryon 
\beq
\B\B=\epsilon^{I_1\cdots I_\N }\epsilon^{J_1\cdots J_\N }\S_{I_1 J_1}\cdots \S_{I_\N J_\N }.\eeq
\end{itemize}
For odd $\N $ the di-baryon is stable, as no baryon exists. 
For even $\N$ it might be co-stable, depending on its binding energy.
{Furthermore, the theory is symmetric under $\SU(\N)$ charge conjugation, and the lightest
C-odd state are glue-balls as $d^{abc}\G_{\mu\mu'}^a\G_{\nu\nu'}^b\G_{\rho\rho'}^c$}.

\subsection{A symmetric of SU($\N $): dynamical symmetry breaking}
The most generic vacuum expectation value can be written as
$\med{S}=\diag(w_1,\ldots,w_\N )$ with $w_i\ge 0$.
Dynamical symmetry breaking is induced by loop corrections, such that we
must take into account the one-loop potential, given by
\beq V_1 = \frac{1}{4(4\pi)^2} \left[
3 \Tr M_V^4 \ln \frac{M_V^2}{\bar\mu^2} + \Tr  M_S^4 \ln \frac{M_S^2}{\bar\mu^2}\right]\eeq
in terms of generic scalar and vector mass matrices.
If $\S$ is written as a vector with real components
with gauge generators $T^A$, the vector mass matrix is
$(M_V^2)^{AB} = \g^2 \frac12 {\S}^T\cdot \{T^A,T^B\}\cdot {\S}$. 
In the limit of small quartics the one-loop potential is dominated by gauge corrections.

\begin{figure}
$$\includegraphics[width=0.45\textwidth]{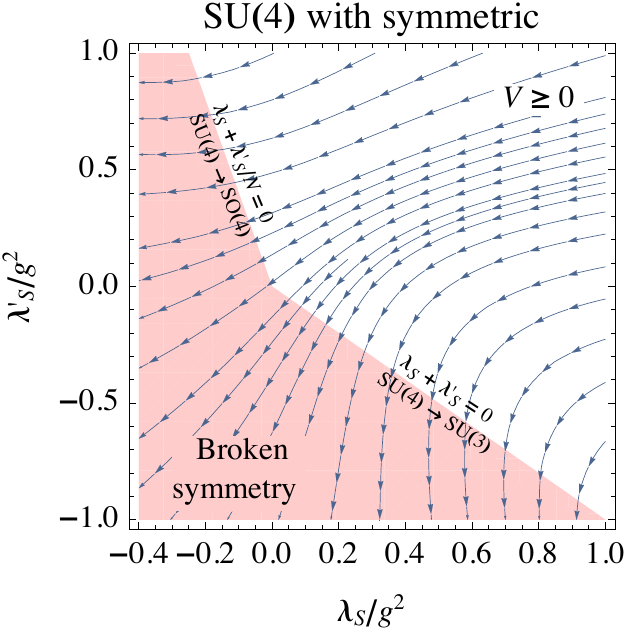}\qquad
\includegraphics[width=0.45\textwidth]{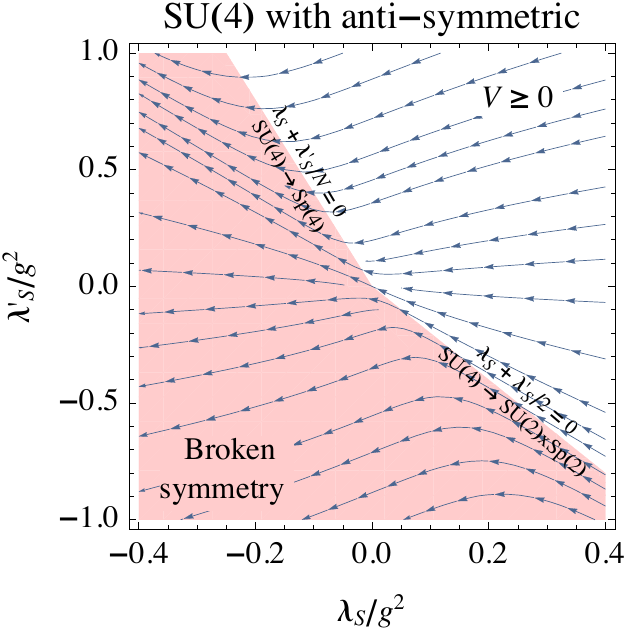}$$
\vspace{-1cm}
\caption{\em \label{fig:flowSU4Sym} 
Coleman-Weinberg symmetry-breaking patterns 
for a $\SU(\N )$ gauge theory with a scalar in the symmetric (left)
and anti-symmetric (right):
the RGE flow towards low energy of its quartics $\lambda_\S$ and $\lambda'_\S$
can intersect the instability conditions in both their branches,
leading to the two different breaking patterns discussed in the text.}
\end{figure}

We only need to consider the one-loop correction along the flat direction $V=0$
that arises at special values of the couplings.
The tree-level quartic potential of $\S$ satisfies $V\ge 0$ when the quartic couplings satisfy
$\lambda_{\S} + \alpha \lambda'_{\S} \ge 0$
for $\alpha=1$ and for $\alpha=1/\N $,
which are the extremal values of 
$\alpha = \Tr(\S\S^\dagger \S\S^\dagger) /\Tr(\S\S^\dagger)^2$.\footnote{Adding the Higgs $H$, 
the condition $V\ge0$ implies the extra conditions
$\lambda_{H} > 0$ and 
$4\lambda_{H} (\lambda_{\S} + \alpha \lambda'_{\S}) > \lambda_{H\S}^2 $.}
Thereby the two possible breaking patterns are:
\begin{itemize}
\item $\med{S}=\diag(0,\ldots,w)$ 
which breaks $\SU(\N )\to\SU(\N -1)$.
The symmetry breaking boundary is
$\lambda_\S+\lambda'_\S=0$. 

\item $\med{S}=\diag(w,\ldots,w)$, which breaks $\SU(\N )\otimes{\rm U}(1)\to\SO(\N )$.
The symmetry breaking boundary is $\lambda_\S+\lambda'_\S/\N =0$. 
\end{itemize}
Fig.\fig{flowSU4Sym} shows that RGE running can cross either boundaries, 
so that both symmetry breaking patterns can be realised dynamically.
Along the special RGE trajectory where both quartics simultaneously cross 0 
we find the breaking $\SU(\N)\to\SO(\N )$.
Dynamical symmetry breaking with a scalar in the symmetric had been studied in~\cite{Gildener:1976ih,Gildener:1975cj}.

\subsection{A symmetric that breaks $\SU(\N)\to\SU(\N -1)$}
We here consider the range of potential parameters such that the scalar acquires
vacuum expectation value $w$ as
\beq \S=
\left(\begin{array}{cccc}
\tilde\S_{11} & \cdots & \tilde\S_{1,\N -1} & 0\cr
\vdots & \ddots & \vdots & \vdots\cr
\tilde\S_{1,\N -1} & \cdots & \tilde\S_{\N -1,\N -1} & 0 \cr
0 &\cdots & 0 &w+s/\sqrt{2}
\end{array}\right)
\eeq
which  breaks the gauge group $\SU(\N )\to\SU(\N -1)$. 
This is the same breaking produced by a scalar in the fundamental,
so it is useful to consider the differences with respect to the model studied in~\cite{Buttazzo:2019iwr}.
As in the model with $\S$ in the fundamental
a global accidental symmetry remains unbroken,
corresponding to the generator $\N \diag( 1, \ldots, 1,0)/(\N -1)$.
Writing the gauge bosons as 
\beq 
T^a \G^a_\mu = 
\left(
\begin{array}{c|c}
	\A_\mu  & \W_\mu / \sqrt{2} \\ \hline
	\W^{*}_\mu /\sqrt{2}  & 0 \\
\end{array}
\right) \,-\, \Z_\mu \sqrt{\frac{\N-1}{2\N}} \left(
\begin{array}{c|c}
	- \I/(\N-1)  & 0  \\ \hline
	0& 1 \\
\end{array}
\right),
\eeq the perturbative spectrum is:
\begin{itemize}
\item $\N (\N -2)$ massless vectors $\A$ in the adjoint of $\SU(\N -1)$
which inherit $\SU(\N)$ transformations under charge conjugation;
\item $(\N -1)$ complex vectors $\W$  in the fundamental of $\SU(\N -1)$ 
and with dark baryon  number $\pm1$
that acquire tree-level mass $M_\W^2 = \g^2 w^2$
by `eating' the corresponding Goldstones scalars with squared mass $m^2=2 (\lambda_\S+\lambda'_\S)w^2$ (massless at  the symmetry breaking point)
{and that transform as $(i\W)\to(i\W)^*$ under charge conjugation};


\item $1$ {C-odd} vector $\Z$ neutral under $\H$ and under dark baryon number
that acquires tree-level mass mass $M_\Z^2 = 4\g^2 w^2(1-1/\N )$
by `eating' the corresponding Goldstone scalar with squared mass $m^2=2 (\lambda_\S+\lambda'_\S)w^2$ (massless at  the symmetry breaking point);

\item the {C-even} scalon $s$ with loop-level mass $M_s$
(its tree-level squared mass $M_s^2 =3m^2$ vanishes at  the symmetry breaking point);

\item the new particle (not present in models where $\S$ fills a fundamental)
is $\N (\N -1)/2$  scalars with squared mass $M^2_{\tilde S}=2w^2\lambda_\S$
that fill a symmetric $\tilde \S$ under $\SU(\N -1)$, with dark baryon number 2
and gauge interactions
\begin{eqnarray}
\Tr|\D_\mu \S|^2 &=& \Tr\bigg|\bigg(\tilde\D_\mu-\frac{2i\g}{\sqrt{2\N (\N -1)}}\Z_\mu\bigg) \tilde\S\bigg|^2 + \frac12(\partial_\mu s)^2 + \frac{\g^2(\N -1)}{\N } \tilde{s}^2\Z_\mu^2 +\\
&&+
\sqrt{2}\g^2 \tilde s \Re (\tilde{\S}^*_{ij} \W_{i\mu}\W_{j\mu})+\frac{\g^2}{2}\W^*_{i\mu}\W_{i\mu}\tilde{s}^2+\g^2\W^*_{i\mu}\W_{j\mu}\tilde\S_{ik}\tilde\S_{jk}^*
\eea
in terms of $\tilde s=s+\sqrt{2}w$ and of the $\H$-covariant $\tilde\D$ derivative. 
They transform as $\tS\to\tS^*$ under charge conjugation.

\end{itemize}
At the symmetry breaking boundary
$\lambda_\S+\lambda'_\S=0$ the Goldstones and the scalon $s$
become massless, and other scalars acquire positive squared masses.
The tree-level potential equals $V = (\lambda_S+\lambda'_\S)s^4/4$ along the scalon
(its second derivative equals $M_s^2$);
thereby at loop level the scalon acquires squared mass
\beq M_s^2 = 2w^2( \beta_{\lambda_\S}+\beta_{\lambda'_\S})=
\frac{2w^2}{(4\pi)^2} \bigg[3 \g^4 \frac{8-16\N +7\N ^2+\N ^3}{\N ^2}+\lambda_S^2(2\N ^2-2\N)\bigg].\eeq
The same result is obtained using the one-loop effective potential~\cite{Gildener:1976ih}.
%
$M_Z^2/M_W^2$ is twice higher than what obtained from a scalar in the fundamental \cite{Buttazzo:2019iwr}.
Apart from the different masses, the new feature is the presence of $\tS$.

$\W$ is a stable DM candidate, and $\tS$ is stable too if $M_\tS <2 M_\W$.
The $\Z$ decays at tree-level to $\tilde \S \tilde \S^*$ and thereby to SM particles.

For $\N >2$ condensation of $\SU(\N -1)$ forms dark baryons
\beq \epsilon \W^{\N -1},\qquad
\epsilon \W^{\N -3}\A\tS,\qquad
\epsilon \W^{\N -5}(\A\tS)^2,\qquad\ldots 
\eeq
as well as dark di-baryons
\beq  \epsilon\epsilon\tS^{\N -1},\qquad \epsilon\epsilon\tS^{\N -2}\W^2,\qquad
\epsilon\epsilon\tS^{\N -3}\W^4,\qquad \ldots .
\eeq
The DM candidate(s) changes depending on $M_\tS/M_\W$.
At the constituent level, if $M_\tS>2M_\W$ the lightest baryon is $\W^{\N -1}$ and the lightest di-baryon is $\W^{2(\N -1)}$:
the di-baryon might be  stable if its binding energy is large enough.
If $M_\tS<2M_\W$, the lightest di-baryon is $\tS^{\N -1}$; for even $\N $ the lightest baryon is $\W \tS^{\N /2 - 1}$, heavier than half of the mass of $\tS^{\N -1}$, so that they are both stable. Instead, for odd $\N $ the lightest baryon is $\tS^{\sfrac{(\N -1)}{2}}$ and the stability of the di-baryon depends again on the binding energies.
Two baryons can merge into one di-baryon: their relative abundance
is usually given by thermal equilibrium at decoupling.
C-odd glue-balls analogous to those of $\SU(\N)$ are stable DM candidates too.
%

%

\subsection{A symmetric that breaks $\SU(\N)\to\SO(\N)$}\label{SU->SO}
We consider again the model of eq.\eq{LagSym}.
An interesting feature is that dark baryon number is spontaneously broken,
but some particles are kept stable by an accidental 
$\mathbb{Z}_2$ symmetry, that arises from charge conjugation as follows.
To define a charge conjugation symmetry of the Lagrangian
that acts on a scalar as
$\S\to \S^*$ one needs to also transform the $\SU(\N )$ gauge-covariant derivative 
$\D_\mu=\partial_\mu + i \G^A_\mu T^A$
as $\D_\mu\to \D^*_\mu$.
In the usual Gell-Mann basis, some $\SU(\N )$ generators associated to some
vectors $\G^a_{\rm real}$
are real and symmetric
(for example, $\sigma_{1,3}$ for $\SU(2)$) and the remaining generators
associated to vectors $\G^b_{\rm imag}$
are purely imaginary and anti-symmetric
(for example, $\sigma_2$ for $\SU(2)$). The needed transformation thereby is
\beq \label{eq:CSU}
\G^a_{\rm real}\to -\G^a_{\rm real}, \qquad \G^b_{\rm imag} \to  \G^b_{\rm imag}.\eeq
This is an automorphism of the gauge group\footnote{The $\SU(\N )$ structure constants vanish, $f_{abc} = 0$, when $a$ corresponds to a real generator, and $b,c$ to imaginary generators, or when a, b, and c all correspond to real generators.} (it is the outer automorphism of $\SU(\N )$ for $\N >2$) and thus a symmetry of the full Lagrangian of eq.\eq{LagSym}.
This charge conjugation symmetry remains unbroken when the symmetric $\S$ acquires 
vacuum expectation value $w$ as
\beq 
\label{eq:Ssymm}
{\S}=\bigg[\bigg(w+\frac{\s}{\sqrt{2\N }}\bigg)\diag(1,\ldots ,1) + (\ts^b + i \tilde\a^b) T^b_{\rm real}\bigg]e^{i {\a}/{\sqrt{2\N }w}}
\eeq
where $\s,\a,\ts^b,\tilde{\a}^b$ are canonically normalized fields. 

After symmetry breaking,
the $\SU(\N)$ vectors $\G$ decompose into $\A$ and $\W$ as follows.
The imaginary generators give rise to massless vectors
in the adjoint (anti-symmetric) of $\SO(\N )$, $\A^a=\G^a_{\rm imag}$.
The real generators give rise to  massive vectors 
(eating the $\tilde\a^b$) in the 
trace-less symmetric of $\SO(\N )$, $\W^b=\G^b_{\rm real}$.

Even after symmetry breaking, the Lagrangian respects
the charge conjugation $\mathbb{Z}_2$ symmetry, under which 
the massive vectors $\W$ and the scalar $\a$ are odd,
and all other fields are even:
\beq \label{eq:CSU2SO}
\s \stackrel{\cal{C}}\to \s, \qquad
\ts \stackrel{\cal{C}}\to \ts, \qquad
\a \stackrel{\cal{C}}\to -\a, \qquad
\A \stackrel{\cal{C}}\to \A, \qquad 
\W \stackrel{\cal{C}}\to -\W.\eeq
At tree level the spectrum is:
\begin{itemize}
\item $\N (\N -1)/2$ massless vectors $\A^a$ in the adjoint of $\SO(\N )$;
\item $\N (\N +1)/2-1$ vectors $\W^b$ in the traceless symmetric of $\SO(\N )$
that acquire mass $M_\W^2 = 4 \g^2 w^2$
by ``eating''  the $\N (\N +1)/2-1$ scalar Goldstones $\tilde  \a^b$;

\item the Goldstone of global accidental U(1), $\a$,
with squared mass $m^{2}=2(\N \lambda_{\S} +  \lambda'_{\S})w^2$;

\item $\N (\N +1)/2-1$  scalars $\ts^b$ in the traceless symmetric of $\SO(\N )$
with squared mass $M^{2}_{\ts}=2 (\N \lambda_\S+3\lambda'_\S)w^2$;
\item a scalar scalon $\s$ with squared mass $M^{2}_{\s}=3m^2 $.
\end{itemize}
At the symmetry breaking boundary
$\lambda_\S+\lambda'_\S/\N =0$ the Goldstones and the scalon $\s$
become massless, and other scalars have positive squared masses.
The tree-level potential equals $V = (\lambda_S+\lambda'_\S/\N )\s^4/4$ along the scalon
(its second derivative is $M_s^2$):
thereby at loop level the scalon acquires squared mass
\beq M_\s^2 = 2\N w^2 (\beta_{\lambda_\S}+ \frac{\beta_{\lambda'_\S}}{\N })=
2\N w^2 \bigg[12 \g^4 \frac{\N ^2+\N -2}{\N ^2}+\lambda_S^2(4\N ^2+4\N -8)\bigg]
.\eeq
The $\a$ is a Goldstone of global accidental U(1): it remains massless
if U(1) is an exact symmetry.\footnote{In the presence of extra massive fermions charged under the 
group $\G$ and carrying a global U(1) charge anomalous under $\G$, 
$\a$ behaves as an axion in the dark sector. 
By introducing extra fermions also charged under QCD,  
one can arrange for a QCD axion with accidental Peccei-Quinn symmetry 
protected up to effective operators of canonical dimension $\N $ \cite{1704.01122}, 
corresponding to $\det \S$. 
For $w \approx 10^{10}$ GeV, this mechanism 
provides a solution of 
the so-called Peccei-Quinn quality problem \cite{Barr:1992qq,Holman:1992us,Kamionkowski:1992mf} as long as $\N \gtrsim 10$.} 
Expanding the Lagrangian, it acquires the following schematic form 
\bea\label{eq:LSUSOschematic}
\Lag&\sim & (\D\A)^2+(\D\W)^2+ (\D \ts)^2+
(\partial \s)^2  +
\frac{(\partial \a)^2}{w^2}[ (w+\s)^2 +\ts^2]+\g^2 \W^2 (w+\s+\ts)^2+ \nonumber\\
&&+
\g \frac{\partial \a}{ w} (w+\s +\ts) \ts \W+
(\lambda_S+\lambda'_S)[ (w+\s)^2+\ts^2 ]^2+ \lambda'_S w \ts^3
\eea
where  order one factors have been omitted (including them the mass term of $\s$ cancels out)
and where
$\D = \partial + i \g\A$ is here the $\SO(\N)$ covariant derivative.
The lightest among the $\cal{C}$-odd states $\W$ and $\a$ is stable.

Let us assume that $\a$ is massless or very light.
Then $\s$ decays into $\a\a$;
$\ts$ into $\A\A$ (at loop level thanks to the $\lambda'_S w \ts^3$ interaction);
$\W$ decays into $\a\ts$ or $\a\A\A$, depending on the
mass hierarchies.\footnote{If $M_\W > M_{\ts}$ this happens at tree level 
via the vertex $\g(\partial \a / w) w  \ts \W$. Otherwise, the decay proceeds 
at one-loop, e.g.~$\W\to \a(\ts\to \A^2)$, or via a kinetic mixing beween $\partial \a$ and $\W$ 
due to the operator $\g(\partial \a / w) \ts^2 \W$. }

At lower energy the pure gauge $\SO(\N )$ confines, without affecting the neutral state $\a$.
SO gauge dynamics respects O-parity~\cite{Buttazzo:2019iwr} which, as discussed in section~\ref{acc}, extends to U-parity in the full $\SU(\N )$ theory. The vacuum expectation values of $\S_{ij}$ preserves U-parity.
Thereby:

\begin{itemize}
\item For $\N $ even the odd glue-ball  
$\B \sim \A^{i_1 i_2} \cdots \A^{i_{\N-1} i_{\N}}\epsilon_{i_1\cdots i_{\N}}$
is stable thanks to 
U-parity 
($\B$ is U-parity odd, so it cannot decay into $\a$'s which are 
U-parity even)
This DM candidate has power-suppressed interactions to SM particles and is accompanied
by dark radiation $\a$.

\item For $\N $ odd one can form glue-balls $\A\A$,
which can decay into SM particles as well as
 into $\a\a$, as they have the same $\cal{C}$-parity and U-parity.
\end{itemize}
This just means that the heavier $\SU(\N)$ dynamics gives 
no qualitatively new effect beyond pure $\SO(\N)$ gauge dynamics
apart from leaving the massless scalar $\a$.

After confinement of $\SO(\N)$, for even $\N$ the stable state are $\a$, odd under $\cal{C}$, and the U-ball $\epsilon \A^{\N/2}$, odd under U-parity. Bound states made with $\W$, e.g.~$\Tr(\A\W\A)$, decay into $\a$ and possibly the U-ball. For odd $\N$, no U-odd state exists, so that the only stable state is $\a$.

The massless Goldstone $\a$ can become a massive pseudo-Goldstone
DM candidate if extra interactions break the spontaneously broken
accidental global U(1)  symmetry while preserving the $\cal{C}$-parity.
An interesting possibility arises for $\N=3,4$:
the potential admits an extra cubic or quartic coupling
$ \lambda''_\S \, \det\S$ that breaks U(1).
The phase of $\S$ can always be redefined such that the extra coupling 
is real, respecting the $\cal{C}$-parity.
The stable pseudo-Goldstone boson $\a$ acquires a squared mass $m_\a^2 \sim \lambda''_\S w^{\N-2} $.
The spin-independent cross section for direct detection of $\a$ dark matter is
suppressed by its possibly small mass,\footnote{The direct detection cross section for massive pseudo-Goldstone boson DM vanishes at tree level~\cite{Gross:2017dan}
for a quadratic U(1) breaking term.
This does not happen for the case considered here.}
\beq
\sigma_{\textrm{SI}} \sim \frac{m_N^4 m_\a^2}{4 \pi v^2 w^2} f^2 \sin^2 (2 \gamma) \left( \frac 1{M_{1}^2}-  \frac 1{M_{2}^2} \right)^2
\eeq
where $f\simeq 0.3$,
$\gamma$ is the mixing angle that diagonalises the Higgs-scalon mass matrix,
$M_{i}$ are the resulting mass eigenvalues.
If $\a$ becomes very massive, DM is the $\C$-odd state $\Tr(\A\W\A)$ if $\N$ is odd, while for even $\N$ both the state $\Tr(\A\W\A)$, odd under $\C$, and the U-ball $\epsilon\A^{\N/2}$, odd under U-parity, are stable.

If the gauge group $\SU(\N)$ is extended to ${\rm U}(\N)$, as in $D$-brane models, 
the massless Goldstone $\a$ is `eaten' by a massive $\Z$ vector.


%

\subsection{Dualities between the confined/Higgs phases?}\label{sec:SUduality}
Some theories (among which $\SU(\N)$ with a scalar in the fundamental)
exhibit a non-trivial feature~\cite{Buttazzo:2019iwr}: the 
confined phase and the Higgs phase give rise to the same 
asymptotic states. 

We now search for possible dualities 
between the purely confined phase
and the two Higgs phases. 
We find that for $\N $ even
the confined phase could be dual to the $\SU(\N )\to \SU(\N -1)$ Higgs phase,  
since there is a map between the spectra and the accidental symmetries in the two phases, 
while for $\N$ odd such correspondence is not possible, 
and hence there should be a phase transition 
when going to strong coupling $\g \sim 4\pi$. 
Analogously, we find a different spectrum in the $\SU(\N) \to \SO(\N)$ Higgs phase.

It should be noted that there could be in principle multiple, physically inequivalent, confined phases, where the various accidental symmetries are unbroken or not.  Analogously, when the residual gauge group in the Higgs phases confines, several sub-phases are possible if condensates breaking the residual global symmetries form.
Lattice studies could help in clarifying these issues. As discussed in section~\ref{VWlike}, we assume here that condensates that break the accidental symmetries do not form.

Let us now consider in turn the various cases: 
\subsubsection*{$\SU(\N )\to \SU(\N -1)$ with $\N$ even}
Baryons in the confined phase are built with the building block $(\G\S)_{IJ}$.
After decomposing $I=\{i,\N\}$, with $i$ spanning over $\SU(\N-1)$, $(\G\S)_{IJ}$ 
contains the two-index combinations $(\A\tS)_{ij}$, $(\Z\tS)_{ij}$, $(\W\W)_{ij}$ and the one-index combinations $(\A\W)_{i}$, $(\Z\W)_{i}$, $\W_{i}$, $(s\W)_{i}$: within each set all operators have the same charge under the global U$(1)$ symmetry,
and the combination of two one-index operator has the same U(1) charge as the two-index ones. Therefore, after mapping the 
two accidental U$(1)$'s (in the broken and unbroken phase) into each other, the baryon of the confined phase maps into a single physical state with the same charge as in the Higgs phase. 
For instance, the baryon $\epsilon (\G\S)^\N $ corresponds to the operators $\epsilon \W^{\N -1}$, $\epsilon \W^{\N -3}(\A\S)$, etc., all with the same quantum numbers. 
To see this, notice that the $\N $-dimensional $\epsilon$ forces exactly one index to be the last one
($I=\N$):
$(\G\S)_{\N J}$ maps into $\W_j$ (or other one-index combinations); 
and $(\G\S)_{i J}$ map into the two-index combinations. 
All operators obtained in this way have the same quantum numbers. 
The same argument holds true for di-baryons, with building block $\S_{IJ}$, that contains $w,s,\tS_{ij},\W_i$ and the charge matches again the number of indices.
 For $\N $ even the map proceeds also in the opposite direction: when the vacuum expectation value gets smaller and smaller there is no distinction between the different operators interpolating the baryon; in particular the one that contains only one one-index operator, say $\epsilon_{i \cdots}\W_i \cdots $, maps into $\epsilon_{\N i \cdots}(\G\S)_{\N i} \cdots$. 
 Furthermore, the C-odd glue-balls in the confined phase are dual to those in the Higgs phase.

\subsubsection*{$\SU(\N )\to \SU(\N -1)$ with $\N$ odd}
For $\N $ odd the map does not exist: in the Higgs phase there are baryons, but not in the confined phase\footnote{More precisely, it is possible to write a composite operator transforming as a {\it fundamental} of $\SU(\N)$, therefore not corresponding to a physical state in the confined phase: in the Higgsed phase, when $\S$ takes a vev, one component of this operator transforms as a singlet of $\SU(\N-1)$ and corresponds to the baryon.}. Therefore, the Higgs phase cannot be dual to the confined one.

\subsubsection*{$\SU(\N )\to \SO(\N)$}
For even $\N $, the baryons in the confined phase are built with the building-block $(\G\S)_{IJ}$.
In the Higgs phase $(\G\S)_{IJ}$ contains, among other combinations, the unbroken gauge vectors $\A_{ij}$ (when 
the $\med{\S}$ part of $\S$ is taken) 
and the combination $(\A \W)_{ij}$ (when the Goldstone part of $\S$ is taken). 
These two combinations have different quantum numbers under charge conjugation
$\cal C$ defined in eq.\eq{CSU2SO}. 
So baryons with different 
$\cal C$-parity quantum numbers can be built in the Higgs phase, 
namely $\epsilon \A^{\N /2}$ and $\epsilon (\A\W) \A^{\N /2-1} $, respectively even and odd under 
$\cal C$.
Thereby the baryon $\epsilon (\G\S)^{\N /2}$ in the confined phase would correspond to two \emph{distinct} baryons in the Higgs phase. For odd $\N $, the di-baryon $\epsilon\epsilon\S^\N $ in the confined phase is stable (protected by the accidental $\U(1)$), whereas the di-baryons in the Higgs phase ($\epsilon\epsilon\A^{\N }$,$\epsilon\epsilon\W^{\N }$,$\epsilon\epsilon\ts^{\N }$,\ldots) are not (since they are not protected by the residual U-parity, relic of the accidental $\U(1)$), and the two phases cannot be dual to each other. 
The C-odd glue-balls in the confined phase have no dual states in the Higgs phase.


\section{An anti-symmetric of SU($\N $)}\label{SUNantisym}
We now consider a scalar $\S_{IJ}$ in the anti-symmetric representation of SU($\N $),
that is generically complex.
For $\N \ge 5$ the Lagrangian is analogous to eq.\eq{LagSym} 
\be\label{eq:LagAntiSym}
\La=\La_{\rm SM}-\frac{1}{4} \mathcal{G}^A_{\mu\nu}\mathcal{G}^{A\,\mu\nu}+\frac14
\Tr (\D_\mu\S)(\D^\mu\S)^\dag -V_\S\ee
with $V_\S$ as in eq.\eq{VS} respecting a global accidental U(1).
For $\N =2$ the anti-symmetric is equivalent to a singlet, and a reality condition can be imposed.
For $\N =3$ the anti-symmetric is equivalent to an anti-fundamental, reducing to the
model with a single quartic already studied in~\cite{Buttazzo:2019iwr}.
The extra Pfaffian invariant ${\rm Pf}\,\S =\sqrt{\det\S}$ is renormalizable for
$\N=\{4,6,8\}$ and breaks the global U(1).
For $\N =4$ it is a mass term and one can impose a self-duality reality condition,
obtaining a real 6 of $\SO(6)=\SU(4)$ already studied in~\cite{Buttazzo:2019iwr}.
For $\N=6$ it gives the extra cubic coupling
$\epsilon^{I_1 J_1 I_2 J_2 I_3 J_3 } \S_{I_1J_1}\S_{I_2J_2}\S_{I_3J_3}$.
For $\N=8$ it gives the extra quartic coupling
$\epsilon^{I_1 J_1 I_2 J_2 I_3 J_3 I_4 J_4} \S_{I_1J_1}\S_{I_2J_2}\S_{I_3J_3}\S_{I_4J_4}$.

\subsection{An anti-symmetric of SU($\N $): confined phase}\label{SUNanticonfined}
The discussion is the same as for the symmetric of $\SU(\N)$ (section~\ref{sec:SUNsymconf}):
an accidental U(1) baryon number is conserved, and baryons must be formed using constituents
with two indices.
The only difference is that, with an anti-symmetric, extra gluons are not needed
to avoid vanishing index contractions: the baryon is now
\beq
\B=\epsilon^{i_1 j_1\cdots i_{n} j_{n}}  \S_{i_1j_1}\cdots 
\S_{i_{n}j_{n}}\eeq
for $\N =2n$ even. For $\N$ odd an extra derivative or gluon is needed to avoid the vanishing due to $\det \S=0$ for the di-baryons. 
The theory is symmetric under charge conjugation, and the lightest C-odd state are 
$d^{abc}\G_{\mu\mu'}^a\G_{\nu\nu'}^b\G_{\rho\rho'}^c$ glue-balls.

\subsection{An anti-symmetric of SU($\N$): dynamical symmetry breaking}
The  most general vacuum expectation value of one anti-symmetric
can be written 
in terms of $\epsilon$ (the $2\times 2$ anti-symmetric tensor) as
\be
\med{\mathcal{S}} =  \begin{cases}
\diag (w_1 \epsilon, \ldots, w_{k} \epsilon) & \hbox{for $\N =2n$ even}\\
\diag (w_1 \epsilon, \ldots, w_{k} \epsilon,0)& \hbox{for $\N =2n+1$ odd}
\end{cases} \label{eq:vev_anti}
\ee
with $w_i\ge 0$.

The most generic breaking patterns can be described as follows.
If all $w_i$ are non-vanishing and different the breaking pattern is $\SU(\N )\to  \SU(2)^n$.
If one $w$ vanishes and $\N $ is odd, the corresponding $\SU(2)$ extends to $\SU(3)$.
If two $w$ vanish, their $\SU(2)^2$ extends to $\SU(4)$ for $\N $ even 
(to $\SU(5)$ for $\N $ is odd).
If two $w$ are equal their $\SU(2)^2$ extends to $\Sp(4)$.
If $k$ of the $w$'s are equal and the remaining $w$ are vanishing, the unbroken
gauge group is $ \SU(\N -2k)\otimes \Sp(2k)$, as written in table~\ref{tab:breakpatterns}.


We next describe the vacuum expectation values realised in renormalizable
quantum field theories.
A quartic potential, depending on the value of its couplings,
has minima corresponding to~\cite{Li:1973mq}:
\begin{itemize}
\item[1)] to minimal $k=1$, namely only one non-vanishing $w\neq0$.
The unbroken gauge group is $\SU(\N -2)\otimes \SU(2)$,
with $\SU(2)=\Sp(2)$.
\item[2)] to maximal $k$, namely all $w$ are equal.
The unbroken gauge group is $\Sp(\tilde \N )$ 
where $\tilde \N =\N $ for $\N $ even, and $\tilde \N =\N -1$ for $\N $ odd.
\end{itemize}
These same two possibilities are realised with Coleman-Weinberg symmetry breaking,
because  they are encountered at the positivity border 
of the tree-level quartic potential of $\S$.
The condition  $V\ge 0$ implies
$\lambda_{\S} + \alpha \lambda'_{\S} \ge 0$
for $\alpha=1/2$ (a single $w$ non-vanishing)
and for $\alpha=1/\tilde{\N }$ (all $w$ equal).
Fig.\fig{flowSU4Sym}b shows that both these conditions can be crossed by the RGE flow,
given for $\N \ge 5$ by
{\small
\begin{eqnsystem}{sys:RGEsym}
(4\pi)^2 \frac{d\g}{d\ln\mu} & = & \frac{2+21\N }{6} \g^3\\
(4\pi)^2 \frac{d\lambda_\S}{d\ln\mu} &=&\g^4\left(\frac{9}{16}+\frac{3}{2\N ^2}\right)
-12 \g^2 \lambda _\S
\left(\N -1-\frac2\N \right)+\\ \nonumber
&&
+96 \lambda^{\prime 2} _S+128(\N -1)   \lambda _\S\lambda' _\S
+32 \left(\N ^2-\N +8\right) \lambda _S^2\\
   (4\pi)^2 \frac{d\lambda'_\S}{d\ln\mu} &=&
   \g^4 \left(\frac{3 \N }{16}-\frac{3}{4}-\frac{3}{\N }\right)-12 \g^2  \lambda' _\S
   \left(\N -1-\frac2\N \right)
   +64  \lambda^{\prime 2} _\S (\N -\frac52)+384 \lambda' _\S \lambda _\S.
\end{eqnsystem}}
We next study the physics that occurs in the two possible breakings.

\subsection{An anti-symmetric  that breaks  SU($\N $) to Sp($\tilde \N $)}\label{sec:SUSpanti}
In section~\ref{SUeven->Sp} we study even $\N =\tilde\N$;
in section~\ref{SUodd->Sp} we study odd $\N = \tilde\N+1$.

\subsubsection{Even $\N $, $\SU(\N)\to\Sp(\N)$}\label{SUeven->Sp}
Loosely speaking, SO is the real part of SU and Sp is its imaginary part.
While the SO invariant tensor is the unit matrix
(giving rise to simple expressions),
Sp is the group of  rotations $U= \exp(i \theta^A T^A)$ that leave
$\gamma_\N=\I_{\N/2} \,\otimes \, \epsilon$ 
invariant, $U^T \gamma_\N U = \gamma_\N $.
So the Sp Hermitian generators $T^A$ satisfy
$ T^{A*}=- \gamma_\N T^A \gamma_\N^{-1} $.
The Sp fundamental is pseudo-real with $\N$ complex components:
$ \gamma_\N  \N^*$ transforms as the fundamental $\N$. 
The adjoint is the real symmetric with dimension $\N(\N+1)/2$.
The anti-symmetric $\ts$ 
has dimension $\N(\N-1)/2-1$, as it
satisfies a reality condition 
\beq \label{eq:realityantisymmSp}
\ts^* = - \gamma_\N \cdot\ts\cdot \gamma_\N  \eeq
and a `trace-less' condition $\Tr(\ts \gamma_\N)=0$
with the Sp invariant tensor $\gamma_\N$.

For even $\N =\tilde \N $ the vectors $\G$ of SU($\N $) form a
massless adjoint (symmetric)
$\A^b$ of Sp($\N $) and a
massive $\W^a$ in the `trace-less' anti-symmetric of Sp($\N $).
The complex anti-symmetric scalar $\S$ of $\SU(\N)$
($\N(\N-1)$ real components) 
decomposes in two singlets $\s$ and $\a$ and two anti-symmetric $\ts$ and $\ta$
under $\Sp(\N)$
($\ta$ gets eaten by the $\W$ vectors).
The embedding of $\Sp(\N)$ in $\SU(\N)$ is non-trivial and can be performed as follows.
The $\SU(\N)$ generators $T^A$ that are $\Sp(\N)$ generators $T^a$
are those such that $\gamma_\N T^A$ is symmetric,
while the broken generators 
are those such that $\gamma_\N T^A$ is anti-symmetric.
Such traceless Hermitian matrices $\widetilde T$ satisfy the
`SU $-$ Sp'  condition
\begin{equation}\label{eq:realityT}
\widetilde{T}^* = - \gamma_\N \cdot \widetilde{T} \cdot \gamma_\N .
\end{equation}
One of such matrices is $\I_\N$, the others satisfy the 
trace-lessness condition $\tr(\widetilde{T}^{\tilde a})=0$.\footnote{For $\N$ even, an explicit orthonormal basis 
for the $\SU / \Sp$ coset matrices $\widetilde{T}^{\tilde a}$ is given by
\beq
\label{eq:TSUSpcoset}
\frac{1}{\sqrt{2}}\( T_{\rm asym}  \otimes \sigma_k \) , \qquad \frac{1}{\sqrt{2}}\( T_{\rm sym}  \otimes \I_2\) ,
\eeq 
where $T_{\rm asym} = T_{\rm imag}$ and $T_{\rm sym} =  \{ T_{\rm real}, \I_{\N/2} / \sqrt{\N} \}$ 
are respectively the antisymmetric and symmetric 
generators of $\U{(\N/2)}$. 
The basis for $\N$ odd is obtained by including the matrices in eq.~\eqref{eq:TSUSpcoset} restricted to the $\N-1$ 
subspace plus those belonging to the $\SU(\N)/\SU(\N-1)$ coset. \label{foot:genTtilde}} 
The anti-symmetric $\S$ of $\SU(\N)$ is expanded in $\Sp(\N)$ multiplets as
\beq 
\label{eq:Santisymm}
{\S}=\bigg[\bigg(w+\frac{\s}{\sqrt{\N /2}}\bigg)\gamma_\N +
2 \sum_{\tilde{a}} (\ts^{\tilde a} + i \tilde\a^{\tilde{a}}) \widetilde{T}^{\tilde a} \cdot \gamma_\N \bigg]e^{i {\a}/{\sqrt{\N /2}w}}
\eeq
such that $\s$ and $\a$ are singlets and
 $\ts$ and $\ta$ form two different 
real anti-symmetric `trace-less'
irreducible representations of $\Sp(\N)$
(as they independently satisfy the reality condition of eq.\eq{realityantisymmSp},
by virtue of $\widetilde{T}\cdot \gamma_\N=(\gamma_\N \cdot \widetilde{T})^*$,
the ordering is important).

The accidental global U(1) is broken by the vacuum expectation value and not
present in the low energy Sp dynamics.

The perturbative spectrum in the broken phase is:
\begin{itemize}
 \item $\N (\N +1)/2 $
 massless vectors $\A$ in the adjoint of $\Sp(\N )$.

 \item $\N(\N-1)/2 -1$  vectors $\W^a$ in the
real `trace-less' anti-symmetric of $\Sp(\N )$ that acquire  mass $M^2_\W={\g^2w^2}$
 by `eating' the $\ta$ 
 with Coleman-Weinberg squared mass $M_\ta^2 =8w^2(\tilde\N\lambda_\S+\lambda'_\S)$
 that vanishes at the symmetry-breaking boundary. 

\item The scalon $\s$.
With Coleman-Weinberg $\SU(\N)\to\Sp(\N)$ breaking its tree-level
squared mass $M_\s^2 = 3 M_\ta^2$
vanishes at the breaking boundary.

\item The massless Goldstone boson $\a$ (unless the global U(1) symmetry is explicitly broken,
for example for $\N=6,8$ by the extra renormalizable term ${\rm Pf}\S$).

 \item $\N(\N-1)/2 -1$ scalars $\tilde{s}^a$
 that fill a real `trace-less' anti-symmetric of $\Sp(\N )$. 
Assuming Coleman-Weinberg breaking its tree-level squared mass equals
$M_{\ts}^2= 8w^2(\tilde\N\lambda_\S+3\lambda'_\S)$.
 \end{itemize}

The vacuum expectation value $\langle\S\rangle$ respects the
charge conjugation symmetry  $\S\to \S^*$, $\D_\mu \to \D_\mu^*$ of SU($\N$). Its action in terms of the $\Sp$-covariant fields is more transparent in terms of Hermitian matrix fields $\ts \equiv \ts^{\tilde a} \widetilde{T}^{\tilde a}$, $\ta \equiv \ta^{\tilde a} \widetilde{T}^{\tilde a}$ that transform as the gauge vectors under $\Sp(\N)$. Since charge conjugation is equivalent to $\S \gamma_\N  \to (\S \gamma_\N )^*$, one has
\beq \label{eq:SpCparity}
\s \stackrel{\cal{C}}\to \s, \quad
\a \stackrel{\cal{C}}\to -\a, \quad
\ts \stackrel{\cal{C}}\to - \gamma_\N \ts \gamma_\N, \quad
\ta \stackrel{\cal{C}}\to  \gamma_\N \ta \gamma_\N, \quad
\A \stackrel{\cal{C}}\to - \gamma_\N \A \gamma_\N, \quad
\W \stackrel{\cal{C}}\to  \gamma_\N \W \gamma_\N.
\eeq
The invariant tensor $\gamma_\N$  appears because $\cal C$ raises/lowers indices in $\U(\N)$-covariant notation. 

Expanding the Lagrangian, it acquires the same
schematic form as the Lagrangian obtained in the case of
$\SU(\N)$ broken to $\SO(\N)$ by a symmetric, eq.\eq{LSUSOschematic},
except that $\D = \partial + i \g\A$ is now the $\Sp(\N)$ covariant derivative. 
Its terms indeed respect the $\cal{C}$-parity of eq.\eq{SpCparity}.\footnote{The cubic $\ts^3$ vanishes for $\N=4$ but this is a special case:
$\SU(4)\simeq \SO(6)$ broken by a 6 to
$\Sp(4)\simeq\SO(5)$.}
Thereby at the perturbative level (i.e.~before considering confinement), 
the lightest among $\W$ and $\a$ is stable, and other particles decay
as in the $\SU\to\SO$ case:
if $\a$ is massless or very light $\s$ decays into $\a\a$,
$\ts$ into $\A\A$,
$\W$ into $\a\ts$ or $\a\A\A$.

A big  difference arises at non-perturbative level:
while SO baryons are stable because odd under O-parity,
Sp baryons decay into mesons because the $\epsilon$ tensor can be decomposed as
$\epsilon^{i_1\cdots i_{\N}} = \gamma_\N^{i_1 i_2}\cdots \gamma_\N^{i_{\N-1}i_{\N}} + \hbox{permutations}$~\cite{Witten:1983tx}. 
Mesons $\Tr(\W\W)$ and glue-balls $\tr(\A\A)$, $\tr(\A\A\A)$
are even under $\cal{C}$-parity and decay.
$\Tr (\W \A)$ is identically zero.
$\Tr(\W\A\A)$ and $\a$ are odd under $\cal{C}$-parity:
thereby the lighter state is a stable DM candidate.

If $\a$ acquires a small mass, it can be
pseudo-Goldstone DM with dominant derivative interactions.
Similarly to the $\SU\to\SO$ case, 
for $\N=6,8$ the potential admits an extra cubic or quartic coupling,
${\rm Pf}\,\S$ (with real coefficient) that breaks U(1) giving mass to $\a$
while respecting $\cal{C}$-parity.
If $\a$ becomes very massive, DM is the $\cal{C}$-odd meson containing one $\W$.
If the gauge group $\SU(\N)$ is extended to ${\rm U}(\N)$, as in $D$-brane models, 
the massless Goldstone $\a$ becomes a massive $\Z$ vector.


\subsubsection{Odd $\N $, $\SU(\N)\to\Sp(\N-1)$}\label{SUodd->Sp}
For odd $\N =\tilde \N +1$
the perturbative spectrum in the broken phase\footnote{$\N=5$ is potentially special because $\Sp(4)=\SO(5)$.
However the $\SU(5)\to\Sp(4)$ discussed here corresponds to a different embedding
(such that  $24=10\oplus 5 \oplus2 \times 4 \oplus 1$)
than the $\SU(5)\to\SO(5)$  (such that $24=10\oplus 14$) discussed in section~\ref{SU->SO}.} is as in section~\ref{SUeven->Sp} plus
\begin{itemize}
\item One singlet vector $\Z$ with mass $M_\Z^2 =\g^2 w^2/\N$.
It acquires mass by `eating' the massless Goldstone $\a$.

\item $2\tilde\N$ vectors $\X$ in the fundamental of $\Sp(\tilde\N)$ with mass $M_\X^2 = \g^2 w^2/4$.
They acquire mass by `eating'
$2\tilde\N$ scalars $\x$ in the fundamental of $\Sp(\tilde\N)$ with 
tree-level Coleman-Weinberg mass $M_\x^2 = 8w^2(2\lambda_\S+\lambda'_\S)$
that vanishes at the symmetry breaking point.
\end{itemize}
The charge conjugation symmetry is again unbroken by the vacuum, with the extra states transforming as
\be
\x \to - \x^* , \quad \X \to - \X^* , \quad \Z \to - \Z.
\ee
Then, as long as binding energies are not so big as to compensate the difference in constituents mass, the $\cal{C}$-odd meson  $\Tr(\W \A\A)$ can decay into $\Z$, which is a stable DM candidate. The massless Goldstone has become a massive $\Z$.

In addition, there is also an unbroken global U(1) symmetry, with generator $ \diag (0,\ldots,0,1)$. At the perturbative level, the only charged state is $\X$. Then, after confinement of $\Sp(\N-1)$, the meson $\X^T \gamma_{\tilde \N} \X$ is stable too.


\subsection{An anti-symmetric that breaks $\SU(\N )\to \SU(\N -2)\otimes\SU(2) $}
After dropping the scalars `eaten' by massive vectors,
the scalar anti-symmetric can be expanded in components as
\beq \S=
\left(\begin{array}{cc}
\tilde \S\ &0\cr
0 &  \epsilon ( w+\s  )\end{array}\right)=
\left(\begin{array}{cccc|cc}
0 & \tilde\S_{12} &\cdots &\tilde\S_{1,\N -2} & 0&0\cr
-\tilde\S_{12} & 0 & \cdots & \tilde\S_{2,\N -2} & 0&0 \cr
\vdots & \vdots & \ddots & \vdots&\vdots&\vdots \cr
-\tilde\S_{1,\N -2} & \cdots  &-\tilde\S_{2,\N -2} & 0 & 0\cr \hline
0 &0 & \cdots & 0 & 0 &  w+\s  \cr
0 &0 & \cdots & 0 & -w-\s &  0 
\end{array}\right)
\eeq
such that the $\tilde\S_{ij}$ and $\s$ are canonically normalized.
Writing the gauge bosons as 
	\beq 
	T^a \G^a_\mu = 
	\left(
	\begin{array}{c|c}
	\A_\mu  &  \W_\mu /\sqrt{2} \\ \hline
	 \W^{*}_\mu/\sqrt{2} & \A_\mu' \\
	\end{array}
	\right) \,-\, \frac{\Z_\mu}{2\sqrt{{\N(\N-2)}}} \left(
	\begin{array}{c|c}
	- 2\I_{\N-2}  & 0  \\ \hline
	0& (\N-2)\I_2 \\
	\end{array}
	\right),
	\eeq the perturbative spectrum in the broken phase is:
\begin{itemize}
 \item $(\N -1)(\N -3)$ 
 massless vectors $\A$ in the adjoint of SU($\N -2$) {which inherit their transformation law under charge conjugation from $\SU(\N)$};
\item3 massless vectors $\A'$ in the adjoint of SU($2$) {which inherit their transformation law under charge conjugation from $\SU(\N)$}; 
 \item $4(\N -2)$ massive vectors $\W$ in bi-fundamental $(\N -2,2)\oplus
 (\overline{\N -2},\overline{2})$ of $\SU(\N -2)\otimes\SU(2)$ with mass $M^2_\W=\sfrac{\g^2w^2}{4}$. {
 Under charge conjugation $(i\W)\to(i\W)^*$};  
 \item 1 {C-odd} massive vector $\Z$
 with mass $M^2_\Z=\sfrac{\g^2w^2(\N -2)}{\N }$
 corresponding to the generator
 $\diag(2, \ldots , 2,-(\N -2), -(\N -2))/2\sqrt{\N (\N -2)}$. 
%
\item The {C-even} scalon $\s$.
 \item $(\N-2)(\N-3)$ scalars that fill an anti-symmetric $\tilde{\S}$ of $\SU(\N-2)$ with mass $M_{\tilde{\S}}^2=16 w^2\lambda_{{\S}}$. 
They transform as  $\tS\to\tS^*$ under charge conjugation. 
\end{itemize}
There is an unbroken conserved global U(1) 
with generator proportional to $\diag(1,\ldots,1,0,0)$.
under which  $\W$ has charge 1,  $\tilde{\S}$ has charge 2 and all  other fields are neutral. 
So $\W$ and $\tilde{\S}$ are possible DM candidates.
Gauge interactions give rise to $\W\W \leftrightarrow \tilde\S$ processes:
\begin{eqnarray}\nonumber
	\frac{1}{4}\Tr|\D_\mu \S|^2 &=& \frac{1}{4}\Tr\bigg|\bigg(\tilde\D_\mu-\frac{2i\g}{\sqrt{\N (\N -2)}}\Z_\mu\bigg) \tilde\S\bigg|^2 + \frac12(\partial_\mu \s)^2 + \frac{\g^2(\N -2)}{2\N } \ts^2\Z_\mu^2 +\\
	&&+
	\frac{\g^2}{2}\epsilon_{\alpha\beta} \ts \Re (\tilde{\S}^*_{ij} \W_{i\mu}^\alpha\W_{j\mu}^\beta)+\frac{\g^2}{4}\W^{*\alpha}_{i\mu}\W_{i\mu}^\alpha\ts^2+\frac{\g^2}{4}\W^{*\alpha}_{i\mu}\W_{j\mu}^\alpha\tilde\S_{ik}\tilde\S_{jk}^*
	\eea
where $\ts=\s+w$; $\tilde\D$ is the SU$(\N-2)$-covariant derivative; {$i,j=1,\dots,\N-2$} are SU$(\N-2)$ indices; 
{$\alpha,\beta=1,2$} are SU(2) indices. 

Assuming that massive particles have masses comparable to $M_\W$,
both unbroken groups $\SU(\N-2)$ and $\SU(2)$ confine at
$\Lambda_{\SU(\N-2)} > \Lambda_{\SU(2)}$, where
\beq \Lambda_i \approx M_\W \exp(-\frac{2\pi}{b_i \alpha_{\rm DC} }),\qquad
b_{\SU(\N-2)} = \frac{11}{3}(\N-2),\qquad
b_{\SU(2)} = \frac{22}{3}.\eeq
$\SU(2)$ interactions respect a custodial symmetry, which is however broken by $\SU(\N-2)$ interactions.
After the double confinement the non-perturbative spectrum contains:
\begin{itemize}
\item Glue-balls of $\SU(\N-2)$ and of $\SU(2)$ that decay to SM particles.
\item The  C-odd glue-balls of $\SU(\N-2)$,
stable because any effective operator that connects the 
$\SU(\N-2)$ and $\SU(2)$ sectors must contain a $\SU(2)$ singlet,
and all $\SU(2)$ singlets are C-even.
\item The unstable $\Z$ and $\s$.
\item Baryons made of $\W$ only:
\begin{itemize}
\item
If $\N$ is odd,  $\SU(\N-2)$ confinement cannot form $\SU(2)$ singlets.
Confinements give rise to bi-baryons made of $\W^{2(\N-2)}$ and to their anti-bi-baryons made of $\bar\W^{2(\N-2)}$.
Extra bi-baryons made of $\W^{\N-2} \bar\W^{\N-2}$ decay to glue-balls
through $\W\bar\W$ annihilations.

\item If $\N$ is even, $\SU(\N-2)$ confinement can form $\SU(2)$ singlets.
Confinements give rise to 
baryons made of $\W^{\N-2}$ and to their anti-baryons made of $\bar\W^{\N-2}$.
\end{itemize}

\item Baryons with a $\W_i\W_j$ pair replaced by one $\tS_{ij}$.
The mass difference is $2M_\W - M_\S + {\cal O}(\Lambda_{\SU(2)})$.
\end{itemize}

%

\begin{figure}
$$\includegraphics[width=0.45\textwidth]{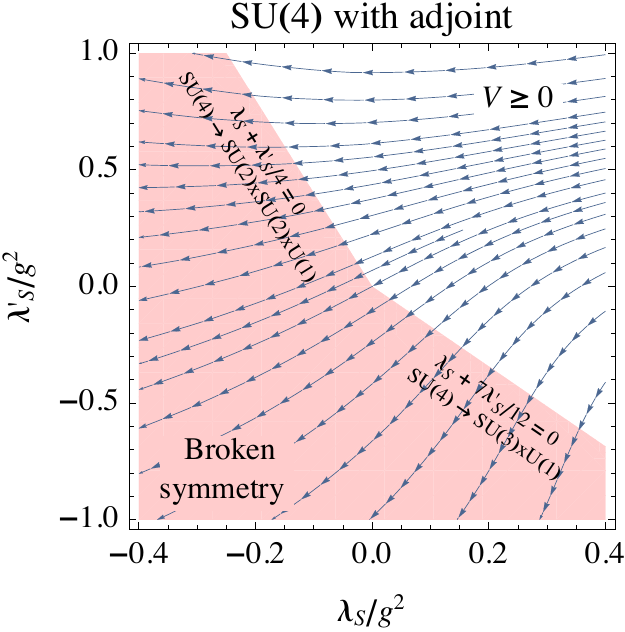}\qquad
\includegraphics[width=0.45\textwidth]{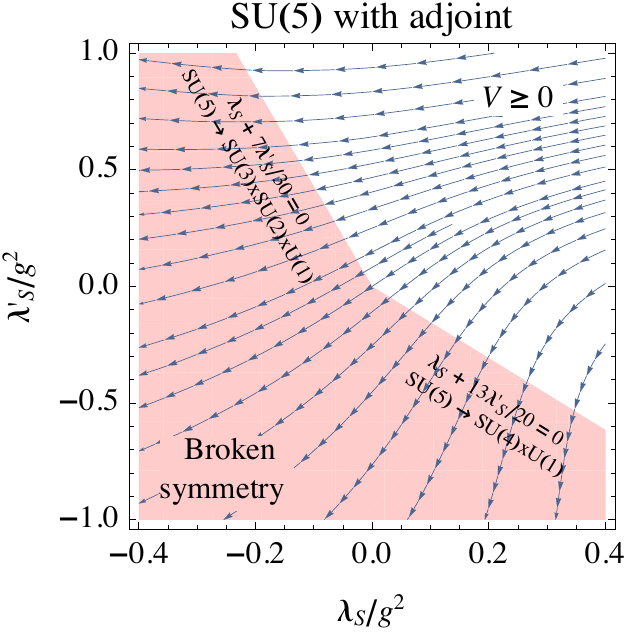}$$
\vspace{-1cm}
\caption{\em \label{fig:flowSU4adj} 
Coleman-Weinberg symmetry-breaking patterns 
for a $\SU(\N )$ gauge theory with a scalar in the adjoint:
the RGE flow towards low energy of its quartics $\lambda_\S$ and $\lambda'_\S$
can intersect the instability conditions in both their branches,
leading to the two different breaking patterns discussed in the text.}
\end{figure}

\section{A trace-less adjoint of SU($\N $)}\label{SUNadj}
Being a real representation, an adjoint $\S^A$ carries no dark baryon number.
Writing it as an Hermitian matrix $\S_I^J = \S^A (T^A)_I^J$ using the
generators in the fundamental,
such that $\Tr\S^2 = (\S^A)^2/2$,
the Lagrangian is
\be\label{eq:Lagadj}
\La=\La_{\rm SM}-\frac{1}{4} \mathcal{G}^A_{\mu\nu}\mathcal{G}^{A\,\mu\nu}+
\Tr (\D_\mu\S)(\D^\mu\S) -V_\S\ee
with
\be V_\S= M_\S^2 \Tr(\S^2)+ A \Tr(\S^3)+ \lambda_\S \Tr (\S^2)^2+
\lambda'_\S \Tr (\S^4)-
 \lambda_{H\S} |H|^2 \Tr \S^2 .
\ee 
for $\N \ge 4$, while $\lambda'_\S$ is redundant for $\N =3$.
Because of the cubic, the theory is not accidentally invariant under 
the $\S\to-\S$ symmetry, so that there are no
stable DM candidates in the perturbative spectrum.
One can consider a dimension-less theory where masses arise dynamically 
as a fermion condensate in a different sector (e.g.\ section 5.1 of~\cite{1705.03896}) 
such that the renormalizable Lagrangian has an accidental $\S\to-\S$ symmetry
and $\S$ only acquires a  mass term through gravitational loops.


\subsection{An adjoint of $\SU(\N)$: confined phase}\label{sec:SUadjconf}
In the presence of the $\S^3$ cubic term there is no stable state. 
One $\S$ decays into two gluons via a loop of $\S$
(the $\S^a \to \G^b \G^c$ amplitude is proportional to the
non-vanishing group-theory factor
$(d_{aij} + i f_{aij}) f_{ikb} f_{jkc}$).
In addition, bound states cannot be stabilized by U-parity (or some other analogous $\mathbb{Z}_2$ symmetry acting non-trivially on the different components), because singlets made with $\G$ and $\S$ necessarily involve indices appearing in pairs: there are no baryons. 
The theory is symmetric under charge conjugation. 
The states odd under C (that acts on $\S$ as $\S\to \S^*$) are $d^{abc}\G ^a \G^b \G^c$, $d^{abc}\G ^a \S^b \S^c$, $f^{abc}\S ^a \G^b \G^c$, $f^{abc}\S ^a \S^b \S^c$
and mix since they have the same quantum numbers. The lightest is a DM candidate.

If cubic terms are absent, the state $\Tr(\S\G)$ becomes stable, with phenomenology similar to the 1-ball of $\SO(\N)$~\cite{Buttazzo:2019iwr}.
The SU(2) case was computed on the lattice~\cite{1906.11193}.

\subsection{An adjoint of $\SU(\N)$: dynamical symmetry breaking}
The RGE for the dimension-less couplings for $\N \ge 3$ are
{\small
\begin{eqnsystem}{sys:RGEadj}
(4\pi)^2 \frac{d\g}{d\ln\mu} & = & -\frac{7\N }{2} \g^3\\
(4\pi)^2 \frac{d\lambda_\S}{d\ln\mu} &=&9\g^4-12\N \g^2 \lambda _\S+
\left(6+\frac{18}{\N^2}\right) \lambda^{\prime 2} _S+8 
   \left(\N -\frac{3}{2\N }\right) \lambda _\S \lambda' _\S+
   2 \left(7 +\N ^2\right) \lambda _\S^2\\
   (4\pi)^2 \frac{d\lambda'_\S}{d\ln\mu} &=&
   3\N\g^4 -12\N \g^2  \lambda' _\S +4  \lambda^{\prime 2} _\S \left(\N -\frac{9}{\N }\right)+
   24 \lambda_\S \lambda'_\S.
\end{eqnsystem}}
The RGE for $\lambda_\S$ for $\N =3$ is obtained replacing
$9$ with $27/2$ and setting $\lambda'_\S=0$.

The tree-level quartic potential of $\S$ satisfies $V\ge 0$ when the quartic couplings satisfy
$\lambda_{\S} + \alpha \lambda'_{\S} \ge 0$,
where $\alpha$ are the extremal values of $\alpha = \Tr(\S\S^\dagger \S\S^\dagger)/\Tr(\S\S^\dagger)^2$. The maximal value is  
$\alpha=(\N^2-3\N+3)/\N(\N-1)$ corresponding to 
$\med{\S} \propto \diag(w,\ldots,w,-(\N -1)w)$, which breaks  
$\SU(\N ) \to \SU(\N -1) \otimes \U(1)$. 
For $\N = 2 k$ even the minimal value is $\alpha=1/\N$ 
corresponding to 
$\med{\S} \propto \diag(w,\ldots,w,-w,\ldots,-w)$, 
which breaks 
$\SU(\N) \to \SU(k) \otimes \SU(k) \otimes \U(1)$. 
For $\N = 2 k +1$ odd the minimal value is 
$\alpha=(\N^2+3)/\N(\N^2-1)$ 
corresponding to 
$\med{\S} \propto \diag(k w,\ldots,k w,k w,-(k+1)w,\ldots,-(k+1)w)$, 
which breaks $\SU(\N) \to \SU(k+1) \otimes \SU(k) \otimes \U(1)$. 
Thereby the possible unbroken gauge groups selected by the Coleman-Weinberg mechanism 
up to e.g.~$\N=6$
are\footnote{The case $\N =3$ is special since $\alpha=1/2$ identically. 
Therefore the quartic scalar potential contains a unique invariant 
$\Tr (\S^2)^2$, such that there is an O$(8)$ accidental 
global symmetry, spontaneously broken to O$(7)$. This yields 7 
Goldstone bosons, of which 4 are eaten by the massive vectors 
of the local $\SU(3) \to \SU(2) \otimes \U(1)$ breaking and 3 remains 
in the physical spectrum. At tree level they are massless, but gauge interactions,
which do not respect the accidental O$(8)$ symmetry of the scalar potential,  
will lift them to $M_{\rm GB} \simeq \g^2 w / 4\pi$.}
\beq \begin{array}{l}
\SU(2) \to {\rm U}(1)\, , \\
\SU(3) \to \SU(2)\otimes {\rm U}(1)\, , \\
\SU(4)\to \SU(3)\otimes {\rm U}(1), ~\SU(2)\otimes\SU(2)\otimes {\rm U}(1)\, , \\
\SU(5)\to \SU(4)\otimes {\rm U}(1), ~\SU(3)\otimes\SU(2)\otimes {\rm U}(1)\, , \\
\SU(6)\to \SU(5)\otimes {\rm U}(1), ~\SU(3)\otimes\SU(3)\otimes {\rm U}(1)\, , 
\end{array}
\eeq
and so on for $\N > 6$. 
In general $\SU(\N_1)\otimes\SU(\N_2)\otimes{\rm U}(1)$ contains a U(1) factor, such that the lightest charged state is a stable DM candidate;
dark photons $\Dg$ are massless; 
as discussed in section~\ref{darkMono}
dark magnetic monopoles exist with
magnetic charge $\g_{\rm mag} = 4 \pi / \g  $ and mass $M_{\rm mag} \sim M_{\W} / \alpha_{\rm DC}$.


%
%
%

\subsection{An adjoint that breaks $\SU(\N)\to \SU(\N-1)\otimes{\rm U}(1)$}
After dropping the scalars `eaten' by massive vectors,
the scalar adjoint can be expanded in block form as
\beq \S=
(w+\s)T^{\N^2-1}+
\left(\begin{array}{c|c}
\tilde \S\ &0  \cr \hline
0 & 0  
\end{array}\right) 
\eeq
where $T^{\N^2-1}=\text{diag}(1,\cdots,1,1-\N)/\sqrt{2\N(\N-1)}$.
Writing the gauge bosons as 
	\beq 
 \G = \begin{pmatrix}
\A & \W^+ \\
\W^-& 0
\end{pmatrix}
	 \,+\, \Dg T^{\N^2-1} ,
	\eeq the perturbative spectrum in the broken phase is: 
\begin{itemize}
\item $(\N-1)^2-1$ massless vectors $\A$ in the adjoint of $\SU(\N-1)$
which inherit their $\SU(\N)$ transformations under charge conjugation.
\item 1  massless C-odd vector $\Dg$ corresponding to the unbroken U(1),
with generator proportional to $\med{\S}$.
\item $\N-1$ complex massive vector $\W^{\pm}$ 
in the fundamental of $\SU(\N-1)$ with mass $M_\W^2= \g^2w^2Q^2$ and with charge $Q=\pm\sqrt{\N/2(\N-1)}$ under the unbroken U(1) gauge group. 
Under charge conjugation $(i\W^+)\to(i\W^+)^*$. 
\item The C-even scalon $\s$ with loop-level mass $M_\s$.
\item $(\N-1)^2-1$ scalars $\tS$ that fill an adjoint of $\SU(\N-1)$ with squared mass $M_\tS^2=w^2(\lambda_{{\S}}+\sfrac{3\lambda_{\S}'}{(\N(\N-1))})$ and neutral under the U(1) gauge group. Under charge conjugation $\tS\to\tS^*$.
\end{itemize}
At perturbative level $\W^\pm$ is a stable DM candidate,
while $\tS$ decays into $\A\A$ through loops involving the $w\tilde{\S}\W\W^*$ gauge coupling.
At non-perturbative level, condensation of $\SU(\N-1)$ gives stable dark baryons $\epsilon\W^{\N-1}$
charged under U(1). 
The states $d^{abc} \A^a \A^b \A^c$, $d^{abc} \A^a \tS^b \tS^c$,   $f^{abc} \tS^a \A^b \A^c$ and $f^{abc} \tS^a \tS^b \tS^c$ are C-odd.
As discussed in the next subsection,
the lightest of such C-odd states
decays into $\Dg$ and C-even states.


\subsubsection*{Glueball decays}
%

The C-even glueballs can decay into dark photons through the dimension-8 interactions
of eq.\eq{dim8}
where the coefficients $C_{1,2} \approx (\N-2)/(4\pi w^{2})^2$, yielding a decay width
\be
\Gamma_{{\rm DG}\to \Dg\Dg} \approx \frac{\alpha_{\rm dark}^4(\N-2)^2}{8\pi}\frac{f_{\rm DG}^2 M_{\rm DG}^3}{(M_\W)^8},
\ee
where $f_{\rm DG} \approx M_{\rm DG}^3$ is a form factor.
This has to be compared with the decay into SM particles, which proceeds through the Higgs-scalon portal. From
\be
\L_{s\A\A} = \frac{\alpha_{\rm dark}}{8\pi}b_\W(\A_{\mu\nu}^a)^2\frac{\s}{w},
\ee
where $b_\W = -7/2$ is the contribution of a loop of $\W$'s to the $\SU(\N-1)$ beta function, one has
\be
\Gamma_{{\rm DG}\to {\rm SM}}\approx \frac{\alpha_{\rm dark}^2 f_{\rm DG}^2 b_\W^2}{512\pi^3}\frac{\lambda_{H\S}^2}{M_{\rm DG} M_{\s}^4}.
\ee
Dark glueball decay predominantly into SM particles if
\be
\left(\frac{M_{\rm DG}}{M_\W}\right)^4\left(\frac{\beta_{\lambda_\S}+\alpha \beta_{\lambda_\S'}}{\alpha b_\W \lambda_{H\S}}\right)^2\left(\frac{8(\N-2)(\N-1)}{\N}\right)^2 \lesssim1, 
~\text{with} ~\alpha=\frac{\N^2-3\N+3}{\N(\N-1)}.
\ee
These results have been used in figure~\ref{fig:Dgplot}.

Assuming that the lightest C-odd state is the glue-ball $d\A\A\A$, it can decay into $\A\A+\Dg$ through the dimension-8 operator $d^{abc}\A^a\A^b\A^c\Dg$, obtained by integrating a loop of $\W$'s. In most of the parameters space the decay is fast. The picture is similar to what discussed in section \ref{darkRadiation} with the difference that the decays of the C-odd glueballs always produce a $\Dg$ in the final state. If the dark photons thermalize with the SM bath after being produced, the contribution to $\Delta N_{\rm eff}$ is given in eq. \eqref{eq:deltaN}. 
In the opposite regime some regions of the parameters space could be excluded by the bound in eq.\eq{DRbound}, depending on the relative amount of C-even/odd glue-balls and their mass ratio. A more detailed analysis is required to determine the precise contribution to $\Delta N_{\rm eff}$. 

\subsection{An adjoint that breaks $\SU(\N)\to \SU(\N_1)\otimes\SU(\N_2)\otimes{\rm U}(1)$}
After dropping the scalars `eaten' by massive vectors,
the scalar adjoint can be expanded in block form as
\begin{equation}
\S = \frac{w+s}{\sqrt{2\N_1\N_2\N}}
\begin{pmatrix}
\N_2 \I_{\N_1} & 0 \\
0 & -\N_1 \I_{\N_2}
\end{pmatrix} \; + \; \begin{pmatrix}
\tS_1 & 0 \\
0 & \tS_2
\end{pmatrix}
\end{equation}
and the gauge vectors as 
\begin{equation}
\G = \begin{pmatrix}
\A_1 & \W^+ \\
\W^-& \A_2
\end{pmatrix} + \frac{\Dg}{\sqrt{2\N_1\N_2\N}}
\begin{pmatrix}
\N_2 \I_{\N_1} & 0 \\
0 & -\N_1 \I_{\N_2}
\end{pmatrix}   .
\end{equation}
The perturbative spectrum is: 
\begin{itemize}
\item 1 massless vector $\Dg$ corresponding to the unbroken U(1).
\item $\N_1^2-1$ massless vectors $\A_1$ in the adjoint of $\SU(\N_1)$, and
$\N_2^2-1$ massless vectors $\A_2$ in the adjoint of $\SU(\N_2)$.
\item $\N_1\N_2$ complex massive vectors $\W^\pm$ in the bi-fundamental of $\SU(\N_1)\otimes \SU(\N_2)$
with squared mass 
$M_\W^2 = \g^2  w^2 Q^2$ and with charge $Q=\pm\sqrt{\N/2\N_1\N_2}$ under the unbroken U(1) gauge group.

\item The scalon $\s$, singlet under the unbroken gauge group.

\item $\N_1^2-1$ neutral real scalars $\tS_1$ in the adjoint of $\SU(\N_1)$
with squared mass $M_{\tS_1}^2 = w^2(\lambda_\S+\sfrac{3\lambda_\S'\N_2}{\N_1\N})$, and
$\N_2^2-1$ neutral real scalars $\tS_2$ in the adjoint of $\SU(\N_2)$ 
with squared mass $M_{\tS_2}^2 = w^2(\lambda_\S+\sfrac{3\lambda_\S'\N_1}{\N_2\N})$.

\end{itemize}
The transformations under charge conjugation are analogous to those of the previous subsection.

The dark scalars $\tS_{1,2}$ in adjoints  of $\SU(\N_{1,2}$)
decay into their gauge bosons in view of their $\tS_{1,2}^3$ cubic couplings.
The massive vectors in the  bi-fundamental $\W^\pm_{i_1 i_2}$ are stable,
for example because they are the lightest state with U(1) charge.

The group with larger $\N_1 > \N_2$ confines earlier,
forming dark baryons $\W^{\N_1}$ in various representations of $\SU(\N_2)$
with $\N_2$-ality $\N_1-\N_2$.
The smallest representation containing the lightest states can be a
singlet if $\N_1=k\N_2$ with integer $k$,
a fundamental if they differ by 1, an anti-symmetric if they differ by 2, etc. 
At lower energy $\SU(\N_2)$ confines forming baryons of baryons.
The lightest DM candidate is charged under U(1), with charge equal to the number of constituents
in this matryoshka.
The abundance of dark photons can be estimated as in the previous section. 

C-odd states such as $d_{{\rm SU}(N_1)}^{abc}\A_1^a\A_1^b\A_1^c$ and $d_{{\rm SU}(N_2)}^{abc}\A_2^a\A_2^b\A_2^c$ (or combination involving scalars) decay into $\Dg$ analogously to the previous subsection.




\section{A trace-less symmetric of $\SO(\N)$}\label{SONsym}
We now consider a scalar $\S_{IJ}$ in the trace-less symmetric real representation of SO($\N $).
The Lagrangian is
\be\label{eq:LagSymSO}
\La=\La_{\rm SM}-\frac{1}{4} \mathcal{G}^A_{\mu\nu}\mathcal{G}^{A\,\mu\nu}+\frac12
\Tr (\D_\mu\S)(\D^\mu\S) -V_\S\ee
with
\be \label{eq:VSSOsym}
V_\S= \frac{M_\S^2}{2}  \Tr(\S^2)+\lambda_\S (\!\Tr \S^2)^2+ A \Tr(\S^3)+
\lambda'_\S \Tr (\S^4)-
 \lambda_{H\S} |H|^2 \Tr( \S^2) .
\ee 
The only special case is $\N=3$: the $\lambda'_\S$ quartic can be removed using $\Tr(\S^4)=\Tr(\S^2)^2/2$; 
the possible extra cubic $\det\S$ equals $\Tr(\S^3)/3$.
For $\N=4$ the possible extra quartic coupling $\det\S$ can be rewritten in terms of $\lambda_\S$ and $\lambda'_S$
 as $\det\S=\Tr(\S^4)/4-\Tr(\S^2)^2/8$.
The action is invariant under O-parity, that can be written as\footnote{The cubic term $\Tr\S^3$
respects O-parity because each $\S$ has two indices.
This differs from the model where $\S$ if a fundamental where O-parity was
a symmetry because a cubic term was forbidden by gauge invariance.}
\be
{\cal G}_{IJ} \stackrel{\P_{\rm O}}{\to} (-1)^{\delta_{1I}+\delta_{1j}} {\cal G}_{IJ} \qquad \textrm{and} \qquad  \S_{IJ} \stackrel{\P_{\rm O}}\to (-1)^{\delta_{1I}+\delta_{1J}}  {\S}_{IJ}. 
\ee 

\subsection{A symmetric of $\SO(\N)$: confined phase}\label{sec:SOsymconf}
We consider the phase where $\SO(\N)$ confines.
Baryons are odd under  O-parity, so that the lightest baryon  is a stable DM candidate.

For even $\N$ baryons exist, made with the constituents $\G_{IJ}$ and $(\G\S)_{IJ}$. 
As long as $M_\S$ is non negligible compared to the confinement scale,
the lightest baryon is the 0-ball $\epsilon \G^{\N/2}$, odd under O-parity.
The detailed structure of $\S$ (a symmetric or a fundamental) does not impact DM phenomenology.

For odd $\N$ no baryon exists.
Differently from the analogous $\SU(\N)$ model, di-baryons $\epsilon\epsilon\S$ are unstable, 
because even under O-parity. Indeed, for $\SO(\N )$ the product of two $\epsilon$'s 
can be expressed as the sum of products of $\N $ Kronecker  $\delta$'s \cite{Witten:1983tx}. 
So there is no DM candidate.

\subsection{A symmetric of $\SO(\N)$: dynamical symmetry breaking}

The tree-level quartic potential of $\S$ satisfies $V\ge 0$ when the quartic couplings satisfy
$\lambda_{\S} + \alpha \lambda'_{\S} \ge 0$ where $\alpha$
are the minimal and maximal values of $\alpha = \Tr(\S^4) /\Tr(\S^2)^2$.
The maximal value is $\alpha=1 - 3/\N-1/(\N-1)$ 
corresponding to $\S \propto \diag(1,\ldots,1,\N-1)$, which breaks $\SO(\N)\to\SO(\N-1)$.
For $\N=2k$ even the minimal value is $\alpha=1/\N $ 
corresponding to $\S \propto \diag(1,\ldots,1,-1,\ldots ,-1)$, which breaks $\SO(\N)\to\SO(k)^2$.
For $\N=2k+1$ odd the minimal value is $\alpha=(3+\N^2)/(\N^3-\N)$
which corresponds to the $\SO(\N)\to\SO(k+1)\otimes \SO(k)$ breaking.
For $\N=3$ the minimal and maximal values of $\alpha$ coincide, corresponding to the unique breaking $\SO(3)\to\SO(2)$.

The RGE for $\N \ge 4$ are 
\begin{eqnsystem}{sys:RGEsym}
(4\pi)^2 \frac{d\g}{d\ln\mu} & = & -\frac{21\N -46}{3} \g^3\\
(4\pi)^2 \frac{d\lambda_\S}{d\ln\mu} &=&18\g^4 -24 \g^2\N \lambda _\S+\\ \nonumber
&&+12 \left(1 +\frac{6}{\N^2}\right)  \lambda^{\prime 2} _S+8 
   \frac{2N^2+3\N-6}{\N} \lambda _\S \lambda' _\S+4 \left(14 +\N +\N ^2\right) \lambda _\S^2\\
   (4\pi)^2 \frac{d\lambda'_\S}{d\ln\mu} &=&
   6\N\g^4-24\N \g^2  \lambda' _\S+4 \frac{2\N^2+9\N-36}{\N}  \lambda^{\prime 2} _\S +96\N \lambda_\S \lambda'_\S.
\end{eqnsystem}
Fig.\fig{flowSO}a shows that, again, 
the RGE flow can cross both stability conditions
depending on the values of the couplings.

\begin{figure}
$$\includegraphics[width=0.45\textwidth]{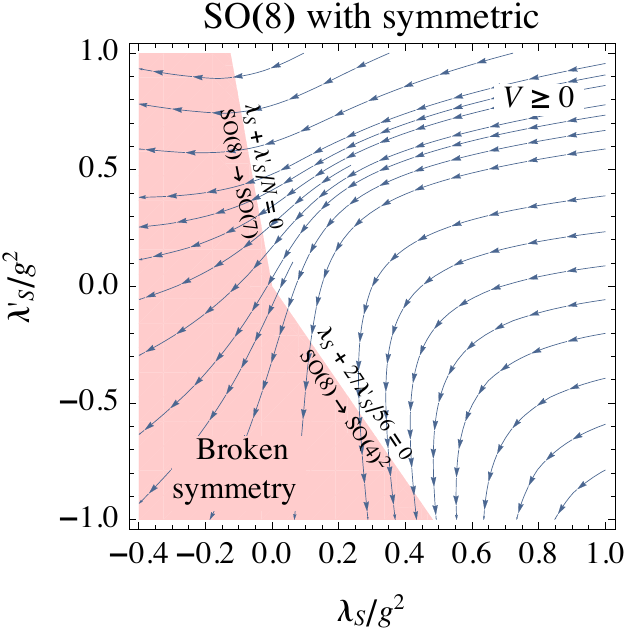}\qquad
\includegraphics[width=0.45\textwidth]{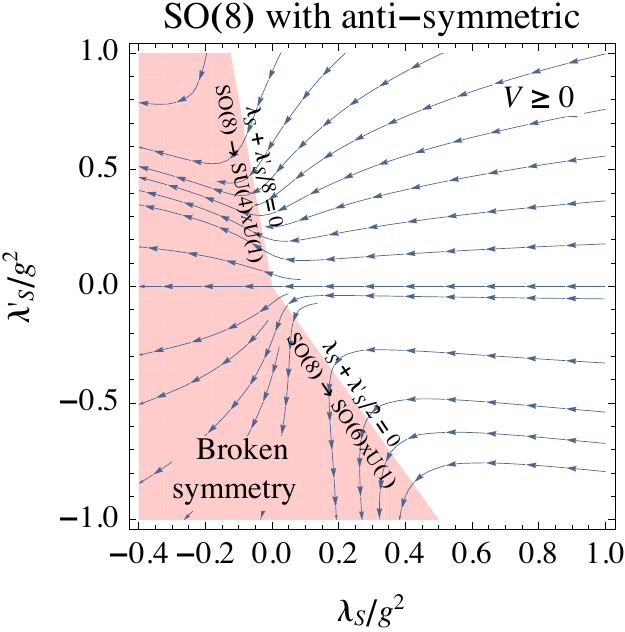}$$
\vspace{-1cm}
\caption{\em \label{fig:flowSO} 
Coleman-Weinberg symmetry-breaking patterns 
for a $\SO(\N )$ gauge theory.
The RGE flow towards low energy of its quartics $\lambda_\S$ and $\lambda'_\S$
can enter both instability conditions,
leading to the two different breaking patterns discussed in the text.}
\end{figure}


\subsection{A symmetric that breaks  $\SO(\N)\to\SO(\N-1)$}
The trace-less symmetric scalar can be  expanded as 
\begin{equation}
\S = \frac{w+s}{\sqrt{\N(\N-1)}} \begin{pmatrix}
\I_{\N-1} & 0 \\
0 & -(\N-1)
\end{pmatrix} \; + \; \begin{pmatrix}
\tS & 0 \\
0 & 0
\end{pmatrix}
\end{equation}
and the gauge vectors as 
\begin{equation}
\G = \begin{pmatrix}
\A & \W \\
-\W^T&0
\end{pmatrix} .
\end{equation}
O-parity is preserved in this vacuum, 
since it can be realised as a reflection along any of the unbroken $\N-1$ directions~\cite{Buttazzo:2019iwr}. 
The perturbative spectrum is: 
\begin{itemize}
\item The scalon $\s$, singlet under $\SO(\N-1)$ and O-parity even.
\item $(\N-1)(\N-2)/2$ massless vectors $\A$ in the adjoint of $\SO(\N-1)$.
\item $\N-1$  vectors $\W$ with squared mass $M_\W^2 =2\N  \g^2 w^2 /(\N-1)$
in the fundamental of $\SO(\N-1)$.
\item $\N(\N-1)/2-1$ scalars in the trace-less symmetric  $\tS$  of $\SO(\N-1)$
with squared mass $M_\tS^2 = 4w^2(\lambda_\S+3\lambda_{\S'}/\N(\N-1))$. 
\end{itemize}
$\tS$ and $\s$ decay into $\A\A$ through loops involving
the $w\tS\W\W$ and $w\s\W\W$ gauge couplings.
The $\W$ is stable: being the only field with one index
it is the only field odd under a $\diag(-1,\dots,-1,1)$ O($\N$) reflection,
respected by the vacuum expectation values.

DM phenomenology remains as in the  
model where $\SO(\N)$ is broken to $\SO(\N-1)$ by a scalar in the fundamental~\cite{Buttazzo:2019iwr},
because $\tS$ is the only extra state.
$\SO(\N-1)$ confines at a scale $\Lambda$ forming  
baryons odd under O-parity and other states.
The lightest baryon is a stable DM candidate.
The baryon  constituents are $\A_{ij}$, $\W_i$. 
For $\N$ even the lightest baryon is the 1-ball $\W\A^{(\N-2)/2}$.
For $\N$ odd the lightest baryon is the 0-ball $\A^{(\N-1)/2}$.


\subsection{A symmetric that breaks $\SO(\N)\to \SO(k)\otimes \SO(\N-k)$}
The trace-less symmetric scalar can be expanded as
\begin{equation}
\S =
\sqrt{\frac{\N-k}{\N k}}(w+s)
\begin{pmatrix}
\I_{k} & 0 \\
0 & - \I_{\N-k} k/(\N-k)
\end{pmatrix} \; + \; \begin{pmatrix}
\tS_1 & 0 \\
0 & \tS_2
\end{pmatrix}
\end{equation}
and the gauge vectors as 
\begin{equation}
\G = \begin{pmatrix}
\A_1 & \W \\
-\W^T& \A_2
\end{pmatrix} .
\end{equation}
Symmetry breaking selects $k=\N/2$ for $\N$ even and $k=(\N - 1)/2$ for $\N$ odd.

O-parity splits into two independent O-parities $\rm O_1$ and $\rm O_2$. This is because reflections acting in the two subspaces are not equivalent under the unbroken gauge group. 
For even $\N$ there is also a twin symmetry that exchanges the two $\SO(\N/2)$. 

The perturbative spectrum is: 
\begin{itemize}
\item The scalon $\s$, singlet under the unbroken gauge group and even under O-parities.
\item $k(k-1)/2$ massless vectors $\A_1$ in the adjoint of $\SO(k)$.
\item $(\N-k)(\N-k-1)/2$ massless vectors $\A_2$ in the adjoint of $\SO(\N-k)$.
\item $k(\N-k)$ vectors $\W$ in the bi-fundamental of $\SO(k)\otimes \SO(\N-k)$
with squared mass 
$M_\W^2 = 2 \g^2  w^2\N/k(\N-k)$.
\item $k(k+1)/2-1$ real scalars $\tS_1$ in the trace-less symmetric of $\SO(k)$ 
with squared mass $M_{\tS_1}^2 = 4w^2(\lambda_\S+\sfrac{3\lambda_\S'(\N-k)}{k\N})$.
\item $(\N-k)(\N-k+1)/2-1$ real scalars $\tS_2$ in the trace-less symmetric of $\SO(\N-k)$ 
with squared mass $M_{\tS_2}^2 = 4w^2(\lambda_\S+\sfrac{3\lambda_\S'k}{\N(\N-k)})$.

\end{itemize}
$\tS_{1,2}$ decay into $\A_{1,2}\A_{1,2}$ through loops involving the $w\tS_{1,2}\W\W$ gauge couplings.
The $\W$ is stable being the only field odd under
the O($\N$) reflection diag($-\I_k,\I_{\N-k}$), which is an unbroken symmetry of
the action.

When the two SO groups confine they can form baryons odd under the two independent O-parities. We need to distinguish 3 cases:
\begin{enumerate}
\item $\N$ even, $\N/2$ even, for example $\SO(8)\to\SO(4)^2$.
In this case the two independently stable 0-balls $\B_1\sim\epsilon \A_1^{k/2}$, $\B_2\sim\epsilon \A_2^{k/2}$ are DM candidates. 
They are degenerate because the entries of their $2\times 2$ mass
matrix is related by the permutation symmetry, 
and because its off-diagonal elements vanish.
The two O-parities are independent: in the full theory the unbroken $\SO(\N)$ O-parity forbids
operators of the form $\B_1\B_2$ and as a consequence  there are no $\B_1\leftrightarrow\B_2$ oscillations.

\item $\N$ even, $k=\N/2$ odd, for example $\SO(6)\to \SO(3)^2$.
Baryons do not exists. However, the bi-baryon 
$\epsilon \epsilon \W \A_1^{(k-1)/2} \A_2^{(k-1)/2}$ is stable since it is odd under both O-parities and no separately odd states exist.

\item $\N$ odd, for example $\SO(7)\to\SO(4)\otimes\SO(3)$.
Without loss of generality, we take $k=(\N\pm 1)/2$ even, $\N-k$ odd. The 0-ball $\epsilon \A_1^{k/2}$ exists and is stable, being odd under $\rm O_1$-parity. 
No $\rm O_2$-parity odd baryons exist
(even states decay) and neither bi-baryons, odd under both O-parities.
While the DM phenomenology is analogous to a $\SO(k)$ model,
its cosmological abundance can be affected by the  extra states.
\end{enumerate}
The case $\N=4$ and $\N=5$ are special.

For $\N=4$ the breaking pattern is
$\SO(4)\to\SO(2)_1\otimes\SO(2)_2$. The particles are a massive $\W$ charged under both the two $\SO(2)=\U(1)$, 
the unstable scalon $\s$, degenerate
massive scalars $\tS_i$ with charge $Q_i(\tS_i)=2Q_i(\W)$ under $\U(1)_i$.
The DM candidate is $\W$, accompanied by $\tS_i$ if $M_{\tS_{1,2}}<2M_\W$.
As all interactions are perturbative, the DM abundance is reproduced for DM masses
that can conflict with the bounds of section~\ref{darkCoulomb}.

%

For $\N=5$ the breaking pattern is $\SO(5)\to\SO(3)\otimes\SO(2)$.
The  perturbative spectrum contains the following massive particles: the scalon $\s$, vectors $\W^\pm$ in the 3 of $\SO(3)$ charged under $\U(1)$, 
a neutral scalar $\tS$ in the symmetric traceless of $\SO(3)$ 
a singlet $\tS^{\pm\pm}$ with $\U(1)$  charge $Q(\tS^{\pm\pm})=2Q(\W^\pm)$. 
Even assuming that its decay $\tS^{\pm\pm}\to\W^\pm\W^\pm$ 
is not kinematically forbidden,
confinement of $\SO(3)$ gives various potentially (co)stable DM candidates,
thanks to conservation of dark $\U(1)$ charge and of O-parity: 
the baryons $\B^\pm=\epsilon\W^\pm\A$ and
$\B^{\pm\pm\pm}=\epsilon\W^{\pm}\W^{\pm}\W^{\pm}$, the meson $\M^{\pm\pm}=\W^\pm\W^\pm$.

\section{An anti-symmetric adjoint of SO($\N $)}\label{SONantisym}
We consider a scalar $\S_{IJ}$ in the anti-symmetric representation of SO($\N $),
with $\N(\N-1)/2$ real components.
The Lagrangian is
\be\label{eq:LagASym}
\La=\La_{\rm SM}-\frac{1}{4} \mathcal{G}^A_{\mu\nu}\mathcal{G}^{A\,\mu\nu}-\frac14
\Tr (\D_\mu\S)(\D^\mu\S) -V_\S\ee
with 
\be \label{eq:VSSOsym}
V_\S= -\frac{M_\S^2}{4}  \Tr(\S^2)+\lambda_\S (\!\Tr \S^2)^2+
\lambda'_\S \Tr (\S^4)-
 \lambda_{H\S} |H|^2 \Tr( \S^2) .
\ee 
The cubic $\Tr(\S^3)$ identically vanishes,
and $\det\S$ vanishes for $\N$ odd.
For $\N=2$ $\SO(2)=\U(1)$ and
the anti-symmetric of $\SO(2)$ is the neutral adjoint of U(1).
For $\N=3$ $\SO(3)=\SU(2)$ and
the anti-symmetric of $\SO(3)$ is the adjoint of $\SU(2)$; furthermore
$\Tr(S^4)=\Tr(\S^2)^2/2$ so that the potential contains only one independent quartic.
For $\N=4$ $\SO(4)=\SU(2)^2$ and 
the anti-symmetric splits into two irreducible representations
$\S_{IJ} = \pm \epsilon_{IJKL}\S_{KL}/2$ 
which are the adjoints of $\SU(2)^2$.
For $\N=6$ $\SO(6)=\SU(4)$ and the anti-symmetric of $\SO(6)$ is the adjoint of $\SU(4)$:
its extra  cubic invariant is written in SO language  as $ {\rm Pf}\,\S$.
For $\N=8$ there is  an extra real quartic invariant $\lambda''_\S{\rm Pf}\,\S$.
The Pfaffian is proportional to
the square root of the determinant of an anti-symmetric matrix with even dimension.
So for $N\neq 6$ the theory has an accidental $\S\to-\S$ symmetry.

\subsection{An anti-symmetric of $\SO(\N)$: confined phase}
The discussion is the same as for the symmetric (section~\ref{sec:SOsymconf}), with in this case baryon constituents $\G_{IJ}$ and $\S_{IJ}$.

\subsection{An anti-symmetric of $\SO(\N)$: dynamical symmetry breaking}
As the anti-symmetric  is the adjoint,  the unbroken group contains a U(1) factor.
In order to understand the unbroken group,
we consider the most generic 
vacuum expectation value of the form of eq.~\eqref{eq:vev_anti} with generic non-vanishing $w_i = w$.
As explained below, we can assume  that $k$ $w_i$ are equal and that the remaining ones vanish.
The breaking pattern is $\SO(\N)\to\SO(\N -2k) \otimes \SU(k) \otimes \U(1)$.
Indeed,
blocks with vanishing vacuum expectation value leave an $\SO(\N - 2k)$ factor unbroken.
Blocks with equal $w=w_i$ form a
$\mathcal{S}_{2 k} =  \I_k \otimes  \epsilon $,
giving rise to a $\SU(k) \times \U(1)$ as follows.
Let us consider generators of the form $S^a \otimes \epsilon$ and $i A^b \otimes \I_2$, 
where $S^a$ and $A^b$ are, respectively, $k \times k$ symmetric and anti-symmetric matrices.
They commute with the vacuum expectation value $\mathcal{S}_{2k}$;
are generators of $\SO(2 k) \subset \SO(\N )$; 
close the $\SU(k) \times \U(1)$ algebra.

Minima of renormalizable potentials~\cite{Li:1973mq} and of Coleman-Weinberg potentials (discussed below)
give rise to vacuum expectation values of the form considered above
with special values of $k=1$ or $k=[\N /2]$, depending on the numerical values
of the quartics.

The tree-level quartic potential of $\S$ satisfies $V\ge 0$ when the quartic couplings satisfy
$\lambda_{\S} + \alpha \lambda'_{\S} \ge 0$ where $\alpha$
are the minimal and maximal values of $\alpha = \Tr(\S^4) /\Tr(\S^2)^2$.
The maximal value is $\alpha=1/2$ 
corresponding to $\S \propto \diag(0,\ldots,0,\epsilon)$, which breaks $\SO(\N)\to\SO(\N-2)\otimes\U(1)$.
For $\N=2k$ even the minimal value is $\alpha=1/\N $ 
corresponding to $\S \propto \diag(\epsilon,\ldots,\epsilon)=\I_k \otimes  \epsilon $, which breaks $\SO(\N)\to\SU(k)\otimes\U(1)$.
For $\N=2k+1$ odd the minimal value is $\alpha=1/(\N-1)$
corresponding to $\S \propto \diag(\epsilon,\ldots,\epsilon,0)$
which corresponds to the $\SO(\N)\to\SU(k)\otimes \U(1)$ breaking.
For $\N=3$ the minimal and maximal values of $\alpha$ coincide, corresponding to the unique breaking $\SO(3)\to\U(1)$.

The RGE for $\N \ge 4$ are
\begin{eqnsystem}{sys:RGEsym}
(4\pi)^2 \frac{d\g}{d\ln\mu} & = & 7(2-\N) \g^3\\
(4\pi)^2 \frac{d\lambda_\S}{d\ln\mu} &=&\frac{9}{2}\g^4 -24 \g^2(\N-2) \lambda _\S+\\ \nonumber
&&+48  \lambda^{\prime 2} _S + 32(2\N-1) \lambda _\S \lambda' _\S+16 \left(16 -\N +\N ^2\right) \lambda _\S^2\\
   (4\pi)^2 \frac{d\lambda'_\S}{d\ln\mu} &=&
   \left(\frac{3}{2}\N-12\right)\g^4-24\left(\N-2\right) \g^2  \lambda' _\S+16\left(2\N-1\right)  \lambda^{\prime 2} _\S +384\N \lambda_\S \lambda'_\S.
\end{eqnsystem}
Fig.\fig{flowSO}b shows that, again, 
the RGE flow can cross both stability conditions
depending on the values of the couplings.

\subsection{An anti-symmetric that breaks SO($\N$) $\to$ SO($\N-2$)$\otimes$ U(1)}
\label{SON->SON-2U(1)}
 	After dropping the scalars `eaten' by massive vectors,
 	the scalar anti-symmetric can be expanded in components as 
\beq \S=
 \left(\begin{array}{cc}
  \tilde\S &0\cr
 0 &  \epsilon ( w+\s  )\end{array}\right)=
 \left(\begin{array}{cccc|cc}
 0 & \tilde\S_{12} &\cdots &\tilde\S_{1,\N -2} & 0&0\cr
 -\tilde\S_{12} & 0 & \cdots & \tilde\S_{2,\N -2} & 0&0 \cr
 \vdots & \vdots & \ddots & \vdots&\vdots&\vdots \cr
 -\tilde\S_{1,\N -2} & \cdots  &-\tilde\S_{2,\N -2} & 0 & 0&0\cr \hline
 0 &0 & \cdots & 0 & 0 &  w+\s  \cr
 0 &0 & \cdots & 0 & -w-\s &  0 
 \end{array}\right)
 \eeq
such that the $\tilde{S}_{ij}$ and $s$ are canonically normalized.
Writing the gauge bosons as \begin{equation}
	\G = \begin{pmatrix}
	\A & \W \\
	-\W^T& 0
	\end{pmatrix} \,+\, \Dg T^{\sfrac{\N(\N-1)}{2}} ,
      \end{equation}
the perturbative spectrum in the broken phase is:
\begin{itemize}
\item The scalon $\s$ with loop-level mass $M_s$.
\item $\sfrac{(N-2)(N-3)}{2}$ massless vectors $\A_\mu$ in the adjoint of SO($\N-2$).
\item 1 massless vector $\Dg$ corresponding to the
unbroken U(1) generator $T^{\sfrac{\N(\N-1)}{2}}$ that performs
rotations of the two latter directions.

\item $(\N-2)$ complex vectors $\W^i_\mu$ with squared mass $M_\W=\g^2w^2 Q^2$ 
 in the real fundamental of SO($\N-2$) and with charge $Q=\pm1$ under the U(1) gauge group. 

\item $\sfrac{(\N-2)(\N-3)}{2}$ real scalars that fill an anti-symmetric $\tilde{\cal S}$ of SO($\N-2$) with mass $M^2_{\tilde{\cal S}}=16w^2\lambda_\S$ neutral under U(1).
\end{itemize}	
The $\W$ is stable because charged under unbroken gauge U(1),
while $\tS$ decays into $\A\A$ through loops involving
the $w\tilde{\S}\W\W^*$ gauge coupling.

We expect  that confinement of $\SO(\N-2)$ at $\Lambda$ does not break the U(1). 
Indeed in the limit $M_\W\gg\Lambda$ the heavy $\W$ form condensates
suppressed by their mass (in analogy with heavy quarks, see e.g.~\cite{1209.0408}).
Strong interactions do not discriminate between
the neutral condensate $ \W_i^+ \W_i^-$
and the charged condensates  $ \W_i^\pm \W_i^\pm$;
weak U(1) interactions favour the neutral condensate $\W^+\W^-$
such that U(1) remains unbroken.
Then U(1) as well as O-parity can lead to stable bound states.
The non-perturbative spectrum contains
\begin{itemize}
\item Unstable states, such as glue-balls and mesons neutral under the U(1),
$\M^0 =\W^+_i \W^-_i$.
\item Mesons possibly stable because charged under the U(1), 
$\M^{\pm\pm}=\W^\pm_i \W^\pm_i$.

\item Baryons odd under O-parity of $\SO(\N-2)$ 
(not broken by the vacuum expectation value) 
and with different U(1) charges:
\beq \begin{array}{lll}
\B^0 \equiv \epsilon \A^{(\N-2)/2},&
\B^{\pm\pm} \equiv \epsilon \W^\pm \W^\pm \A^{(\N-4)/2},\ldots & \hbox{for $\N$ even}\\
\B^{\pm} \equiv \epsilon  \W^\pm\A^{(\N-3)/2},&
\B^{\pm\pm\pm} \equiv \epsilon \W^\pm \W^\pm\W^\pm
 \A^{(\N-5)/2},\ldots & \hbox{for $\N$ odd}
 \end{array}
.\eeq
\end{itemize}
The stable states are as follows:
\begin{itemize}
\item For even $\N$, the lightest baryon $\B^0$ and the lightest
charged state, $\B^{\pm\pm}$ or $\M^{\pm\pm}$
(they are co-stable if both their decay channels
$\B^{\pm\pm} \to \B{}^0 \M^{\pm\pm}$ and
$\M^{\pm\pm} \to\B^0 \B^{\pm\pm} $ are  kinematically closed;
higher-charge baryons could also be co-stable).

%

\item For odd $\N$, the lightest baryon $\B^\pm$.
Likely, $\M^{\pm\pm}$ is co-stable and 
$\B^{\pm\pm\pm}$ decays into  $ \B^\pm \M^{\pm\pm}$.

\end{itemize}

\subsection{An anti-symmetric that breaks $\SO(\N)\to \SU(k) \otimes{\rm  U}(1)$}\label{sec:SOtoSU}
\subsubsection{For even $\N$, $k=\N/2$, $\SO(\N)\to \SU(\sfrac{\N}{2}) \otimes{\U}(1)$} 
After dropping the scalars `eaten' by massive vectors,
the scalar in the anti-symmetric adjoint can be expanded in block form as 
	\beq\label{eq:SofSO2SU}
	\S=
	\sqrt{\frac{1}{k}}\left(w + s\right) \I_{k} \otimes i\sigma_2
	+ i \tS^a {T}^a_{\rm SU}
	\eeq
where $a=1,\cdots,k^2-1$ and ${T}^a_{\rm SU}$ are the $\SU(k)$ generators explicitly given by
	\begin{equation}\label{eq:antiSO}
	\sqrt{2} (T_{\rm SU})_{\rm asym} \otimes \I_2, \qquad 
         \sqrt{2} (T_{\rm SU})_{\rm sym} \otimes \sigma_2 .	
	\end{equation} 
The perturbative spectrum in the broken phase is:
\begin{itemize}
\item 1 massless vector $\Dg$ corresponding to the unbroken U(1) {with generator 
${(\I_{k}\otimes \sigma_2)/\sqrt{k}}$ proportional to $\med{\S}$}. 
\item $k^2-1$ massless vectors $\A$ in the adjoint of SU($k$) {corresponding to the generators of eq.\eq{antiSO}}.
\item $\sfrac{k(k-1)}{2}$ complex massive vectors $\W^{\pm\pm}_{ij}$ with mass $M^2_\W=\g^2w^2 Q_\W^2$ 
that fill an anti-symmetric of SU($k$) and with charge $Q_\W=\pm2/\sqrt{k}$ under the U(1) gauge group. 
They correspond to the combinations
$\W^{\pm\pm} \sim (T_{\rm SU})_{\rm asym} \otimes (\sigma_3\pm i\sigma_1)$. 
\item the scalon $\s$ with loop-level mass $M_s$.
\item $k^2-1$ scalars $\tilde{\S}$ neutral under U(1)
that fill an adjoint of SU($k$) with mass $M^2_{\tilde{\S}}=8w^2(2 \lambda_{\S}+3\lambda_{{\S}}'/k)$.
\end{itemize} 
Analogously to section~\ref{SON->SON-2U(1)},
the $\W$ is stable because charged under unbroken gauge U(1),
while $\tS$ decays into $\A\A$ through loops involving
the $w\tilde{\S}\W\W^*$ gauge coupling.


At non-perturbative level SU confinement acts as in section~\ref{SUNanticonfined},
forming charged baryons $\epsilon\W^{\N/4}$ for  $\N/2$ even, or
di-baryons for $\N/2$ odd. 

The $\SU(\N/2)$ subgroup is invariant under charge conjugation $\C$, whose extension to the full $\SO(\N)$ group we dubbed $\C_{\SO}$. The massless $\Dg$ is C-odd (its generator $\I_{k}\otimes \sigma_2$ is odd under $\C_{\SO}$). Therefore the C-odd glue-balls of $\SU(\N/2)$ decay into $\Dg$ (plus additional C-even states).

\subsubsection{For odd $\N$, $k=(\N-1)/2$, $\SO(\N)\to \SU((\N-1)/2) \otimes{\rm  U}(1)$}
After dropping the scalars `eaten' by massive vectors,
	the scalar in the anti-symmetric adjoint can be expanded in block form as 
	\beq
	\S'=
	\diag\left(\S,0\right) 
	\eeq
where $\S$ is given by eq.\eq{SofSO2SU}.
The perturbative spectrum in the broken phase is as in the previous section, with an extra state:
\begin{itemize}
\item 1 massless vector $\Dg$ corresponding to the unbroken U(1). 
\item $k^2-1$ massless vectors $\A$ in the adjoint of SU($k$).

\item $\sfrac{k(k-1)}{2}$ complex  vectors $\W^{\pm\pm}_{ij}$ with squared mass $M^2_\W=\g^2w^2 Q_\W^2$ that fill an anti-symmetric of SU($k$) and with charge $Q_\W=\pm2/\sqrt{k}$ under the U(1).
 \item the scalon $\s$ with loop-level mass $M_s$.
 \item $k^2-1$ scalars $\tilde{\S}$ that fill an adjoint of SU($k$) 
neutral under the U(1) with mass $M^2_{\tilde{\S}}=8w^2(2 \lambda_{\S}+3\lambda_{{\S}}'/k)$.

\item $k$ extra complex massive vectors $\X^\pm_i$ with mass $M^2_\X=\g^2w^2 Q_\X^2$ that fill a fundamental of SU($k$) and with charge $Q_\X=\pm1/\sqrt{k}$ under the U(1).
\end{itemize}
The $\X^\pm$ is stable, being the lightest state charged under unbroken gauge U(1).
In view of its mass, $\W^{\pm\pm}$ decays into $\X^\pm\X^\pm$;
$\tS$ decays into $\A\A$ through loops involving the $w\tilde{\S}\W\W^*$ gauge coupling.

At non-perturbative level, confinement of $\SU(k)$ leads to stable dark-charged baryons
$\B\sim\epsilon \X^{k}$.

\color{black}

\section{A symmetric adjoint of Sp($\N $)}\label{SpNsym}
We now consider a scalar $\S$ in the 
symmetric adjoint representation of $\Sp(\N )$,
which is a real representation.
Since  the symmetric is the  adjoint,
it can be written in terms of components $\S^A$
($A=1,\ldots,\N(\N+1)/2$) as $\S_I^J=(\S^AT^A)_I^J$
that transforms as $\S\rightarrow U\S U^{\dagger}$
and obeys $\S=\S^\dag$.
$\S$ also satisfies the reality condition
$\S^* = \gamma_\N \S\gamma_\N$,
analogous to eq.\eq{realityantisymmSp},
such that it decomposes into
quaternionic $2\times 2$ blocks~\cite{1711.04656}. 
However $\S$ is not a symmetric matrix,
because the Sp generators $T^A $ are not symmetric.

An equivalent description that makes the symmetry explicit
is obtained lowering one index
obtaining $\hat \S_{IJ} \equiv (\S\gamma_{\N})_{IJ}$
which transforms as $\hat\S\rightarrow U\hat\S U^{T}$.
$\hat\S$ is symmetric (because $T^A \gamma_{\N}$ is symmetric)
and obeys the reality condition 
$\hat\S ^* = \gamma_\N  \hat\S \gamma_\N$.
However $\hat\S$ is not hermitian, $\hat\S\neq\hat\S^\dag$.
We use the $\S$ representation.

The Lagrangian for an Sp adjoint is analogous to the SU adjoint, eq.\eq{Lagadj}:
\be\label{eq:LagSpN}
\La=\La_{\rm SM}-\frac{1}{4} \mathcal{G}^A_{\mu\nu}\mathcal{G}^{A\,\mu\nu}+
\Tr (\D_\mu\S)(\D^\mu\S) -V_\S\ee
except that the cubic $\Tr(\S^3)$ identically vanishes, as well as other odd powers
(as clear using the $\hat{\S}$ representation,
such that one $\gamma_\N$ is needed to contract indices)
such that
\be \label{eq:VSSpN}
V_\S= M_\S^2 \Tr(\S^2)+\lambda_\S (\!\Tr \S^2)^2+
\lambda'_\S \Tr (\S^4)-
\lambda_{H\S} |H|^2 \Tr \S^2.
\ee 
For $\N=2$ $\Sp(2)=\SU(2)$ and $\Tr(\S^4)=\Tr(\S^2)^2/2$
and there is only one quartic.
The renormalizable action is invariant under $\S\to-\S$,
that can be  explicitly broken by non-renormalizable operators,
and spontaneously broken by a vacuum expectation value of $\S$.



\subsection{An adjoint of $\Sp(\N)$: confined phase}
In view of the $\S\to-\S$ symmetry,
the lightest bound state containing an odd number of $\S$ is a stable DM candidate.
Presumably this is the  state $\Tr (\S\G) $. 
Sp baryons decay into mesons, because the $\epsilon$ tensor
can be expanded as combinations of $\gamma_\N$.

\subsection{An adjoint of $\Sp(\N)$: dynamical symmetry breaking}
As the symmetric  is the adjoint,  the unbroken group contains a U(1) factor.

The tree-level quartic potential of $\S$ satisfies $V\ge 0$ when the quartic couplings satisfy
$\lambda_{\S} + \alpha \lambda'_{\S} \ge 0$ where $\alpha$
are the minimal and maximal values of $\alpha = \Tr(\S^4) /\Tr(\S^2)^2$.
The maximal value is $\alpha=1/2$ 
corresponding to
$\S \propto \diag(0,\ldots,0,1,-1)$, which breaks $\Sp(\N)\to\Sp(\N-2)\otimes\U(1)$.
The minimal value is $\alpha=1/\N $ 
corresponding to $\S \propto \diag(1,-1,\ldots,1,-1)$, which breaks $\Sp(\N)\to\SU(\N/2)\otimes\U(1)$.

The RGE for $\N \ge 4$ are
\begin{eqnsystem}{sys:RGEsym}
(4\pi)^2 \frac{d\g}{d\ln\mu} & = & -\frac{7}{4}(\N+2) \g^3\\
(4\pi)^2 \frac{d\lambda_\S}{d\ln\mu} &=&\frac{9}{2}\g^4 -6 \g^2(\N+2) \lambda _\S+\\ \nonumber
&&+3  \lambda^{\prime 2} _S + 2(2\N+1) \lambda _\S \lambda' _\S+ \left(16 +\N +\N ^2\right) \lambda _\S^2\\
   (4\pi)^2 \frac{d\lambda'_\S}{d\ln\mu} &=&
   \frac{3}{2}\left(\N+8\right)\g^4-6\left(\N+2\right) \g^2  \lambda' _\S+\left(2\N+1\right)  \lambda^{\prime 2} _\S +24 \lambda_\S \lambda'_\S.
\end{eqnsystem}
Fig.\fig{flowSp} shows that, again, 
the RGE flow can cross both stability conditions
depending on the values of the couplings.

\subsection{An adjoint that breaks Sp($\N$) $\to$ Sp($\N-2$)$\otimes$ U(1)}
After dropping the scalars `eaten' by massive vectors,
the scalar adjoint can be expanded in block form as
\beq \S=
\left(\begin{array}{c|cc}
\tilde \S\ &0 & 0 \cr \hline
0&(w+\s)/2 & 0 \cr
0&0&-(w+\s)/2 
\end{array}\right) 
\eeq
and the gauge bosons as 
\beq \G\sim
\left(\begin{array}{c|cc}
\A\ &-\gamma_{\N-2} \X^- & \X^+ \cr \hline
\X^{+T} \gamma_{\N-2}&\Dg & \W^{++} \cr
\X^{-T}&\W^{--}&-\Dg 
\end{array}\right) .
\eeq

The perturbative spectrum in the broken phase is: 
\begin{itemize}
\item $\sfrac{(\N-1)(\N-2)}{2}$ massless vectors $\A$ in the adjoint of Sp($\N-2$).
\item 1 massless vector $\Dg$ corresponding to the unbroken U(1),
with generator proportional to $\med{\S}$.
\item 1 complex massive vector $\W^{\pm\pm}$ 
singlet of Sp($\N-2$) with mass $M_\W^2=\g^2w^2 Q_\W^2$ and with charge $Q_\W=\pm1$ under the unbroken U(1) gauge group.
\item $\N-2$ complex massive vectors $\X^\pm$ in the fundamental of Sp($\N-2$) with mass $M_\X^2=\g^2w^2 Q_\X^2$ and with charge $Q_\X=\pm1/2$ under the unbroken U(1) gauge group.
\item The scalon $\s$ with loop-level mass $M_\s$.
\item $\sfrac{(\N-1)(\N-2)}{2}$ scalars $\tS$ that fill a symmetric (adjoint) of $\Sp(\N-2)$ with squared mass $M_\tS^2=w^2\lambda_{{\S}}$ and neutral under the U(1) gauge group.
\end{itemize}
At perturbative level $\X^\pm$ is a stable DM candidate,
$\W^{\pm\pm}$ could be co-stable or decay to
$\X^\pm\X^\pm$ (at tree level $M_\W=2 M_\X$, so that loop corrections
are needed to establish if the decay is kinematically allowed);
$\tS$ decays into $\A\A$ through loops involving the $w\tilde{\S}\X\X^*$ gauge coupling.

We expect that confinement of $\Sp(\N-2)$ leaves U(1) unbroken.
The reason is analogous to section~\ref{SON->SON-2U(1)}:
the possible condensates involve an even number of charged $\X$, 
so that a neutral condensate is possible
and energetically favoured by the weak gauging.
After confinement of $\Sp(\N-2)$ 
the lightest charged states are $\W^{\pm\pm}$ and the charged mesons $\M^{\pm\pm} \equiv \X^{\pm T} \gamma_{\N-2} \X^\pm$. 
At tree level they have the same constituent mass, but
the meson becomes slightly heavier
than the $\W$
so that they could be co-stable.
They co-annihilate such that the non-perturbative annihilation
cross-section of the meson depletes their common abundance.
If the mass splitting  is $\Delta M \sim {\rm keV}$ one gets
the direct-detection phenomenology known as `inelastic DM'.


\subsection{An adjoint that breaks Sp($\N$) $\to$ SU($\sfrac{\N}{2}$)$\otimes$ U(1)}
After dropping the scalars `eaten' by massive vectors,
the scalar adjoint can be expanded in block form as 
\beq
\S=
\sqrt{\frac{1}{2\N}}\left(w + \s\right) \diag(1,-1,\ldots,1,-1)
+ \tS^a {T}^a_{\rm SU}
\eeq
where $a=1,\cdots,(\sfrac{\N}{2})^2-1$ and ${T}^a_{\rm SU}$
are the $\SU(\N/2)$ generators explicitly given by 
\begin{equation}\label{eq:SUinSp}
\frac{1}{\sqrt{2}} (T_{\rm SU})_{\rm asym} \otimes \I_2, \qquad 
\frac{1}{\sqrt{2}} (T_{\rm SU})_{\rm sym} \otimes \sigma_3 .
\end{equation} 
The perturbative spectrum is:
\begin{itemize}
\item $(\sfrac{\N}{2})^2-1$ massless vectors $\A$ in the adjoint of SU($\N/2$),
corresponding to the generators of eq.\eq{SUinSp}.

\item 1 massless vector $\Dg$ corresponding to the unbroken U(1), with generator 
$({\I\otimes \sigma_3})/{\sqrt{2 \N}}$ proportional to $\med{\S}$.

\item $\sfrac{\N(\N+2)}{8}$  complex  massive vectors $\W^\pm$ that fill a symmetric of $\SU(\N/2)$ with squared
mass $M_\W^2=\g^2w^2 Q_\W^2$ and with charge $Q_\W=\pm\sqrt{2/\N}$ under the unbroken U(1). 
They correspond to 
$\W^\pm \sim T_{\rm sym} \otimes \sigma_\pm$, 
where $T_{\rm sym} = \{T_{\rm real}, \I_{\N/2}/\sqrt{\N}\}$ 
are the symmetric generators of $\U(\N/2)$.


\item The scalon $\s$ with loop-level mass $M_\s$.

\item $(\sfrac{\N}{2})^2-1$ scalars that fill a adjoint representation $\tS$
of $\SU(\N/2)$ with squared mass $M_\tS^2 = w^2(\lambda_\S+3\lambda'_{\S}/\N)$
and neutral under the U(1) gauge group.
\end{itemize}
The $\W^\pm$ are stable DM candidates, while the $\tS$ decay into $\A\A$
through the $w\tS \W^+\W^-$ gauge interaction.

When $\SU(\N/2)$ confines, we expect that gauge U(1) remains unbroken.\footnote{As argued in section~\ref{VWlike}, based on Vafa-Witten-like considerations we
expect that no charged $\det(\W)$ condensate forms.
This is surely true for $\N/2$ even
because, even if $\det(\W)$ acquires a condensate,
it is energetically favoured to stay in its neutral component
(while for $\N/2$ odd all components of $\det(\W)$ are charged).
If the gauge U(1) gets broken, its massive vector is a stable DM candidate with exponentially 
suppressed mass proportional to the $\det(\W)$ condensate 
and tiny kinetic mixing $\epsilon \sim \lambda_{H\S}e \g/(4\pi)^4$ with the photon
(achieving such small values naturally is usually difficult). 
Similar considerations can be done whenever there is an unbroken gauged $\U(1)$, e.g. in section~\ref{SUNadj}.}
Then the lightest charged bound state is stable. 
SU condensation proceeds as described in section~\ref{sec:SUNsymconf}, 
with the replacements $\N \to \N/2$, $\G \to \A, \tS$: the stable states are baryons and/or di-baryons, according to the parity of $\N/2$. 

The $\SU(\N/2)$ subgroup is invariant under charge conjugation $\C$, whose extension to the full $\Sp(\N)$ group we dubbed $\C_{\Sp}$. Since the Sp generator $\I_{k}\otimes \sigma_3$ is even under $\C_{\Sp}$, the massless $\Dg$ is C-even.
Therefore the lightest C-odd state, presumably the  C-odd glue-balls of $\SU(\N/2)$, are stable DM candidates.

\begin{figure}
$$\includegraphics[width=0.45\textwidth]{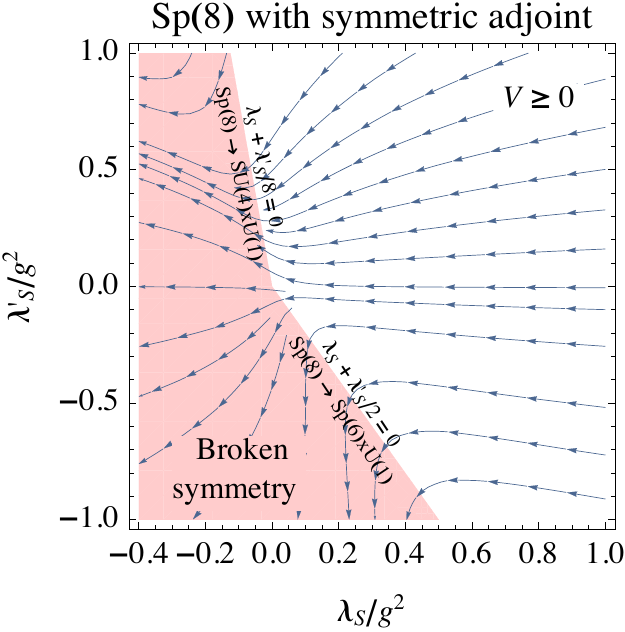}\qquad
\includegraphics[width=0.45\textwidth]{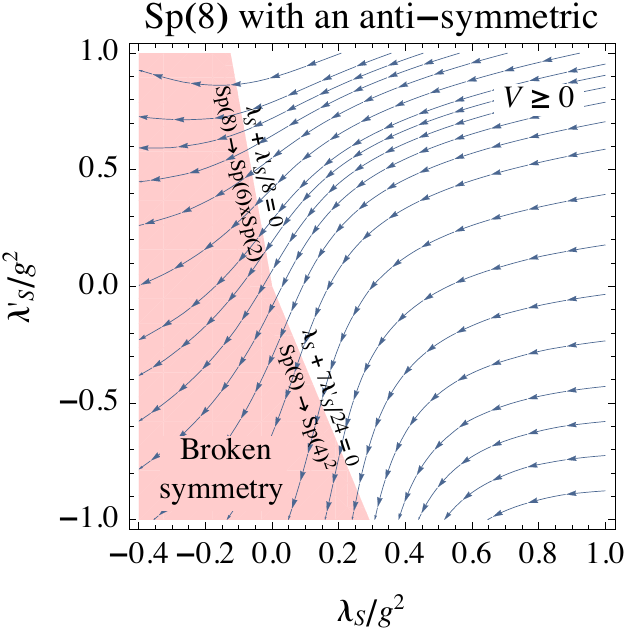}$$
\vspace{-1cm}
\caption{\em \label{fig:flowSp} 
Coleman-Weinberg symmetry-breaking patterns 
for a $\Sp(\N )$ gauge theory.
The RGE flow towards low energy of its quartics $\lambda_\S$ and $\lambda'_\S$
can enter both instability conditions,
leading to the two different breaking patterns discussed in the text.}
\end{figure}

\section{A trace-less anti-symmetric of Sp($\N$)}\label{SpNantisym}
We now consider a scalar $\S$ in the 
anti-symmetric representation of $\Sp(\N)$.
It can be expanded in components $\S^a$ as
\beq
	\S_I^J= (\S^a \tilde{T}^a)_I^J,  \qquad  \hbox{with}\qquad  a=1,\ldots,\N(\N-1)/2-1.
\eeq
where the generators $\tilde{T}^a$ are the sub-set of the Sp generators in the
fundamental that satisfy the `SU $-$ Sp' condition (see eq.~(\ref{eq:realityT}) and footnote \ref{foot:genTtilde}).
The matrix $\S$ transforms as $\S\to U\S U^{\dagger}$ and obeys $\S=\S^{\dagger}$, 
the reality condition $\S^*=-\gamma_N\cdot\S\cdot\gamma_\N$,
and the traceless condition $\Tr(\S \gamma_\N)=0$, 
but it is not an anti-symmetric matrix. 

An equivalent description that makes anti-symmetry manifest is obtained lowering one index obtaining $\hat \S_{IJ} \equiv (\S\gamma_{\N})_{IJ}$ 
($\tilde{T}^a\gamma_{\N}$ is anti-symmetric)
that transforms as $\hat \S\to U\hat \S U^T$.
However $\hat\S$ is not hermitian, $\hat\S\neq\hat\S^\dag$.
We use the $\S$ representation.
	
The Lagrangian is
\be\label{eq:LagSpNAnti}
\La=\La_{\rm SM}-\frac{1}{4} \mathcal{G}^A_{\mu\nu}\mathcal{G}^{A\,\mu\nu}+
\Tr (\D_\mu\S)(\D^\mu\S) -V_\S\ee
with
\be \label{eq:VSSpNAnti}
V_\S= M_\S^2 \Tr(\S^2)+\lambda_\S (\!\Tr \S^2)^2+A\Tr (\S^3)+
\lambda'_\S \Tr (\S^4)-
\lambda_{H\S} |H|^2 \Tr \S^2.
\ee 
For $\N=4$ and $\N=6$ there is only one quartic in view of
$\Tr(\S^4)=\Tr(\S^2)^2/{4}$.
For $\N=4$ the cubic vanishes because the anti-symmetric of
$\Sp(4)$ is the fundamental of $\SO(5)$.
Again, cubics are allowed for fields that satisfy a trace-less condition.
The action is invariant under the $\C_{\Sp}$ accidental symmetry,
which will not lead to stable DM states.

%

\subsection{An anti-symmetric of Sp($\N$): confined phase}
Charge conjugation $\C_{\Sp}$ does not lead to  stable states. 
As in section \ref{sec:SUadjconf}, if cubic terms are absent,
the $\S\to-\S$ symmetry implies that the lightest bound
states made by an odd number of $\S$ is stable. 
As  $\Tr(\G\S)$
vanishes by anti-symmetry, the lightest state
presumably is $\S\S\S$, corresponding to the operator
 $\Tr(\S^3)$.

\subsection{An anti-symmetric of Sp($\N$): dynamical symmetry breaking}\label{Spanti1}
Even in the Sp case, the most generic vacuum expectation value
of an anti-symmetric representation has the form of eq.\eq{vev_anti}~\cite{1711.04656}.

The tree-level quartic potential of $\S$ satisfies $V\ge 0$ when the quartic couplings satisfy
	$\lambda_{\S} + \alpha \lambda'_{\S} \ge 0$ where $\alpha$
	are the minimal and maximal values of $\alpha = \Tr(\S^4) /\Tr(\S^2)^2$.
	The maximal value is $\alpha=\sfrac{(\N^2-6\N+12)}{(2\N^2-4\N)}$ 
	corresponding to
\beq \langle\S\rangle \propto \diag(-2,\ldots,-2,\N-2,\N-2)\eeq
which breaks $\Sp(\N)\to\Sp(\N-2)\otimes\Sp(2)$.
For $\N=4k$ the minimal value is $\alpha=1/\N $ 
corresponding to $\langle\S \rangle\propto (-\I_{2k},\I_{2k})$, which breaks $\Sp(4k)\to\Sp(2k)^2$. For $\N=4k+2$ the minimal value is $\alpha=\sfrac{(\N^2+12)}{\N(\N^2-4)}$ corresponding to $\langle\S \rangle\propto \diag(-{k}\I_{2k+2},({k+1})\I_{2k})$, which breaks $\Sp(4k+2)\to\Sp(2k+2)\otimes\Sp(2k)$. For $\N=4,6$ the minimal and maximal values of $\alpha$ coincide, corresponding to the unique breakings $\Sp(4)\to \Sp(2)\otimes \Sp(2)$ and $\Sp(6)\to \Sp(4)\otimes \Sp(2)$.

The RGE for $\N \ge 8$ are 
\begin{eqnsystem}{sys:RGEantisymSp}
(4\pi)^2 \frac{d\g}{d\ln\mu} & = & -\left(\frac{7}{4}\N +\frac{23}{6} \right)\g^3\\
(4\pi)^2 \frac{d\lambda_\S}{d\ln\mu} &=&\frac{9}{2}\g^4 -6\N \g^2 \lambda _\S+\left(\N^2 -\N+14\right)  \lambda^{2}_S+\\ \nonumber
&&+\frac{4\N^2-6\N-12}{N}\lambda _\S \lambda' _\S+\left(3+\frac{18}{\N^2}\right) \lambda^{\prime 2}_\S\\
(4\pi)^2 \frac{d\lambda'_\S}{d\ln\mu} &=&
\frac{3}{2}\N\g^4-6\N \g^2  \lambda_\S+ 24\lambda_\S \lambda'_\S+ \frac{2\N^2-9\N-36}{N}\lambda^{\prime 2} _\S .
\end{eqnsystem}
Fig.\fig{flowSp}b shows that, again, 
the RGE flow can cross both stability conditions
depending on the values of the couplings.


\subsection{An anti-symmetric that breaks $\Sp(\N)\to \Sp(\N_1)\otimes \Sp(\N_2)$}	\label{Sp1Sp2}
After dropping the scalars `eaten' by massive vectors,
the trace-less anti-symmetric scalar $\S$ can be expanded as:
\beq
\S= \frac{w+\s}{\sqrt{2\N\N_1\N_2}}
\bigg.\left(
\begin{array}{cc}
-\N_2 \I_{\N_1} & 0 \\
0 & \N_1\I_{N_2} \\
\end{array}
\bigg.\right)+
\bigg.\left(
\begin{array}{cc}
\tS_1 & 0 \\
0 & \tS_2 \\
\end{array}
\bigg.\right).
\eeq
with $\N_2=\N-\N_1$.
For $\N=4k$ one has $\N_1=\N_2=\N/2$ e.g.\ 	$\Sp(8)\to\Sp(4)^2$.
For $\N=4k+2$ one has $\N_1=(\N+2)/2$ and $\N_2 = \N_1-2$,
e.g.\ 	$\Sp(6)\to\Sp(4)\otimes\Sp(2)$.
Writing the gauge bosons as 
\begin{equation}
	\G = \begin{pmatrix}
	\A_1 & \W \\
	\W^T& \A_2
	\end{pmatrix} ,
	\end{equation} 
the perturbative spectrum is:
\begin{itemize}
\item $\sfrac{\N_1(\N_1+1)}{2}$ massless gauge bosons $\A_1$ in the adjoint of Sp($\N_1$).
\item $\sfrac{\N_2(\N_2+1)}{2}$ massless gauge bosons $\A_2$ in the adjoint of Sp($\N_2$).
\item $\N_1\N_2$ real massive bosons $\W_{ii'}$ in the bi-fundamental of $\Sp(\N_1)\otimes \Sp(\N_2)$ with mass $M^2_\W=\sfrac{\N\g^2w^2}{2\N_1\N_2}$.
\item the scalon $\s$ with loop-level mass $M_{\s}$.
\item $\sfrac{\N_1(\N_1-1)}{2}-1$ real scalars $\tS_1$ in the trace-less anti-symmetric of Sp($\N_1$) with mass $w^2[\lambda_\S+\sfrac{3\lambda'_{\S}\N_2}{\N\N_1}]$.
\item $\sfrac{\N_2(\N_2-1)}{2}-1$ real scalars $\tS_2$ in the trace-less anti-symmetric of Sp($\N_2$) with mass $w^2[\lambda_\S+\sfrac{3\lambda'_{\S}\N_1}{\N\N_2}]$.
\end{itemize}
For $\N_1=\N_2=\N/2$ the dynamics is invariant under a 
permutation symmetry that exchanges the two Sp($\N/2$) groups.	
For any $\N$ the $\tS_1(\tS_2)$ decays into $\A_1\A_1(\A_2\A_2)$ through loops involving the $w\tS_1(\tS_2)\W\W$ gauge coupling.
The $\W$ is stable because the vacuum and the action of eq.~\eqref{eq:LagSpNAnti} 
leave unbroken a gauge discrete $\mathbb{Z}_2$ symmetry, corresponding  to
the Sp transformation  $\G\to U\G U^{\dagger}$, $\S\to U\S U^{\dagger}$ with $U=(-\I_{\N_1},\I_{\N_2})$.
This $\mathbb{Z}_2$  acts on the fields as
\beq
\s \to \s, \qquad
\tS_1 \to \tS_1, \qquad
\tS_2 \to \tS_2, \qquad
\A_1 \to \A_1, \qquad
\A_2 \to \A_2, \qquad
\W \to -\W.
\eeq
$\W$ would be a stable DM candidates.
But, when the two $\Sp(\N_1)$ and $\Sp(\N_2)$ 
factors in the unbroken sub-group $\H$ confine,
$\W$ form  bi-mesons 
$\M= \gamma_{\N_1}^{i_1j_1} \gamma_{\N_2}^{i_2j_2}\W_{i_1i_2} \W_{j_1 j_2}$.
$\M$ would be stable if the two Sp-arities of the two factors were symmetries of the full theory.
However the full theory respects a unique $\C_{\Sp}$ and
$\M$ decays in view of $\W\W\A_1\A_2$ interactions.
This theory predicts no DM
candidate (unless one considers small enough coupling $\g$ that confinements
happen on super-horizon scales).

\subsection{An anti-symmetric that breaks Sp($\N$) $\to$ Sp($\N-2$)$\otimes$ Sp(2)}\label{SpSp2}
After dropping the scalars `eaten' by massive vectors,
the scalar $\S$ in the anti-symmetric representation can be expanded in block form as 
\begin{equation}
\S =\frac{w+\s}{\sqrt{4\N(\N-2)}} \begin{pmatrix}
-2\I_{\N-2} & 0 \\
0 &(\N-2) \I_{2}
\end{pmatrix} \; + \; \begin{pmatrix}
\tS & 0 \\
0 & 0
\end{pmatrix}
\end{equation}
Writing the gauge bosons as 
\begin{equation}
	\G = \begin{pmatrix}
	\A & \W \\
	\W^T& \A'
	\end{pmatrix} , 
\end{equation}
the perturbative spectrum is: 
\begin{itemize}
\item $\sfrac{(\N-2)(\N-1)}{2}$ massless vectors $\A$ in the adjoint of Sp($\N-2$).

\item 3 massless vector $\A'$ corresponding in the adjoint of unbroken Sp(2).

\item $2(\N-2)$ massive vectors $\W_{i i'}$ that fill a bi-fundamental representation of $\Sp(\N-2)\otimes\Sp(2)$ with squared mass $M^2_\W=\sfrac{\N\g^2w^2}{4(\N-2)}$. 
It satisfies a non trivial  reality condition.

\item The scalon $\s$ with loop-level mass $M_\s$.

\item $\sfrac{(\N-2)(\N-3)}{2}-1$ scalars that fill an anti-symmetric representation of Sp($\N-2$) with squared mass $M_\tS^2 = w^2[\lambda_\S+\sfrac{6\lambda'_{\S}}{\N(\N-2)}]$.
The analogous field for the second $\Sp(2)$ is absent.

\end{itemize}
$\tS$ decays into $\A\A$ through loops involving the $w\tS\W\W$ gauge coupling. 
The $\W$ is stable but forms unstable $\M=\gamma_{2}^{ij} \gamma_{\N-2}^{i'j'} \W_{i i'} \W_{j j'}$ in the confined phase, as already discussed in section~\ref{Sp1Sp2}.


%
%
%
%

\section{Conclusions}\label{concl}
We considered Quantum Field Theory models with one gauge group $\G$ and one scalar $\S$,
exploring the possible groups and two-index representations.
These renormalizable theories often have accidental symmetries  
implying one or more stable particles that are possible Dark Matter candidates.
The accidental symmetries  can be classified as:
\begin{itemize}
\item Global U(1) can arise accidentally when $\S$ lies in a complex (or pseudo-real) representation of $\G$, see section~\ref{General}.

\item Group parities, namely reflections in group space analogous to the usual parity,
act on components of vectors and scalars $\S$, rather than on full multiplets.
Theories with $\G=\SU(\N)$ or $\SO(\N$) often accidentally respect a group parity, unlike $\Sp(\N)$
theories.

\item Group charge conjugations, that do not give extra stable states.
\end{itemize}

Furthermore, extra symmetries (and DM candidates) arise when
$\S$ acquires vacuum expectation values such that $\G$
gets broken to a sub-group that contains a gauge $\mathbb{Z}_2$ discrete symmetry
or a gauge U(1).

In the latter case, the theory contains massless (or exponentially light) dark photons $\Dg$,
and some DM candidates are charged under the gauge U(1), so that DM interacts with $\Dg$.
This gives rise to a specific DM phenomenology, summarized in section~\ref{Dg}:
dark radiation in addition to dark matter;
DM particles elastically scatter with each other and with dark photons;
the possibility of dark monopoles.

\begin{landscape}

\begin{table}
\caption{\small \em Summary of the broken phases. 
For simplicity we loosely include gauge discrete symmetries among accidental global symmetries:
$\P$ denotes a group parity; $\C$ a group charge conjugation.
Possibly stable particles and unbroken local $\U(1)$ are in \cgre{green}.
$\B$ denotes a dark baryon, $\B\B$ a di-baryon,  $\M$ a dark meson.
Theories with low $\N$ can be special. 
\label{tab:summary}}\setlength{\tabcolsep}{3pt}
\vspace{0.5cm}
\begin{small}\hspace{-0.2cm}
\begin{tabular}{|c|cc|cc|cc|c|}
\rowcolor[cmyk]{0,0,0.2,0}
\multicolumn{3}{c}{Unbroken phase}  & \multicolumn{4}{|c|}{Broken phase: perturbative} & \multicolumn{1}{c}{Broken condensed}\\ 
\hline
\rowcolor[cmyk]{0.1,0,0.1,0} gauge& scalar &  accidental & unbroken & accidental  & massive & massive & Dark\\
\rowcolor[cmyk]{0.1,0,0.1,0} $\G$ &rep $\S$ & global &  gauge $\H$ &  global $\H$ & vectors & scalars &  Matter\\
\toprule
\multirow{8}{*}{$\SU(\N)$} &  
fundamental & U(1), $\C$ & $\SU(\N-1)$ & U(1), $\C$ & $\cgre{\W},{\Z}$ & $\s$ & $\B \sim \W^{\N-1}$, $d\A\A\A$ \\
\cline{2-8} 
&\multirow{2}{*}{symmetric}  &   
\multirow{2}{*}{U(1), $\C$} 
& $\SU(\N-1)$ & U(1), $\C$ & $\cgre{\W},{\Z}$ & $\s,\cgre{\tS}$ & $\B \sim \W^{\N-1}$, $d\A\A\A$, 
$\tS$ can be co-stable\\ \cline{4-8}
&&& $\SO(\N)$ & $\P_{\rm U},\C$ & $\W$ & $\cgre{\a},\s,\ts$ & $\a$ + 0-ball if $\N$ even\\  
\cline{2-8}
&\multirow{3}{*}{antisymm}  &
\multirow{3}{*}{U(1), $\C$}  & 
$\SU(\N-2)\otimes\SU(2)$ & U(1),$\C$ &  $\cgre{\W},\Z$ & $\s,\cgre{\tS}$ & $\B\sim\W^{\N-2}$
if $\N$ even; $ \W^{2(\N-2)}$ if odd, $d\A\A\A$
\\ \cline{4-8}
&&& $\Sp(\N)$ & $\C$ &  $\W$ & $\cgre{\a},\s,\ts$ & $\a$\\ \cline{4-8}
&&& $\Sp(\N-1)$ & $\C$, U(1) &  $\W,\cgre{\Z}, \cgre{\X}$ &  $\s,\ts$ & $\Z,\M\sim\X^{T} \gamma_{\N-1} \X$\\ 
\cline{2-8}
&\multirow{2}{*}{adjoint}  &
\multirow2{*}{$\C$}  & 
$\SU(\N_1)\otimes\SU(\N_2)\otimes\cgre{\U(1)}$ & $\C$ &  $\cgre{\W^\pm}$ & $\s,\tS_1,\tS_2$ & 
charged double $\B$, depends  on $\N_{1,2}$\\ \cline{4-8}
&&& $\SU(\N-1)\otimes\cgre{\U(1)}$ & $\C$ &  $\cgre{\W^\pm}$ & $\s,\tS$ & charged $\B\sim \W^{\N-1}$\\ 
\midrule
\multirow{9}{*}{$\SO(\N)$} & 
fundamental & $\P_{\rm O}$ & $\SO(\N-1)$ & $\P_{\rm O}$ & $\cgre{\W}$ & $\s$ & 0(1)-ball for $\N$ odd (even) \\
\cline{2-8} 
& \multirow{4}{*}{\parbox{1.5cm}{\centering symmetric\\[-1.5mm] traceless}}  &  \multirow{4}{*}{$\P_{\rm O}$}
& $\SO(\N-1)$ & $\P_{\rm O}$ &$\cgre{\W}$ & $\s,\tS$ & 0(1)-ball  for $\N$ odd (even)
\\  
\cline{4-8}
&&& $\SO(\N/2)\otimes\SO(\N/2), \N/2$ even & $\P_{{\rm O}_1}$,$\P_{{\rm O}_2}$ &$\cgre{\W}$ & $\s,\tS_1,\tS_2$ & 0-ball \\
\cline{4-8}
&&& $\SO(\N/2)\otimes\SO(\N/2), \N/2$ odd & $\P_{{\rm O}_1}$,$\P_{{\rm O}_2}$ &$\cgre{\W}$ & $\s,\tS_1,\tS_2$ & 1-ball bi-baryon \\
\cline{4-8}
&&& $\SO((\N+1)/2)\otimes\SO((\N-1)/2)$ & $\P_{{\rm O}_1}$,$\P_{{\rm O}_2}$ &$\cgre{\W}$ & $\s,\tS_1,\tS_2$ & 0-ball \\
\cline{2-8}
&\multirow{4}{*}{\parbox{1.5cm}{\centering antisymm\\[-1.5mm] adjoint}}  & \multirow{4}{*}{$\P_{\rm O},\mathbb{Z}_2$} & 
$\SO(\N-2)\otimes\cgre{\U(1)}, \N$ even & $\P_{\rm O}$ &$\cgre{\W^\pm}$ & $\s,\tS$ & neutral
0-ball + charged 2-ball or $\M^{\pm\pm}$ \\
\cline{4-8}  
&&&$\SO(\N-2)\otimes\cgre{\U(1)}, \N$ odd & $\P_{\rm O}$ &$\cgre{\W^\pm}$ & $\s,\tS$ & charged 1-ball $\B^\pm$ + possibly $\M^{\pm\pm}$ \\
\cline{4-8} 
&&& $\SU(\N/2)\otimes\cgre{\U(1)}, \N$ even & -- &$\cgre{\W^{\pm\pm}_{ij}}$ & $\s,\tS$ & charged baryon $\W^{\N/4}$ or dibaryon
\\
\cline{4-8}
&&& $\SU((\N-1)/2)\otimes\cgre{\U(1)}, \N$ odd & -- &$\W^{\pm\pm}_{ij},\cgre{\X^\pm_i}$ & $\s,\tS$ & charged baryon $\X^{(\N-1)/2}$
\\
\midrule
\multirow{5}{*}{$\Sp(\N)$} &
fundamental & U(1) & $\Sp(\N-2)$ & U(1) & $\cgre{\W},\cgre{\X},{\Z}$ & $\s$ & $\W$ and $\M\sim\X^{T} \gamma_{\N-2} \X$ \\
\cline{2-8} 
&  \multirow{2}{*}{\parbox{1.5cm}{\centering symmetric\\[-1.5mm] adjoint}}  & 
\multirow{2}{*}{$\mathbb{Z}_2$} &
$\Sp(\N-2)\otimes\cgre{\U(1)}$ & -- &$\cgre{\W^{\pm\pm}},\cgre{\X^\pm_i}$ & $\s,\tS$ & 
$\W^{\pm\pm}$ and $\M^{\pm\pm}\sim\X^{\pm T} \gamma_{\N-2} \X^\pm$
\\
\cline{4-8} 
&&& $\SU(\N/2)\otimes\cgre{\U(1)}$ & $\C$ &$\cgre{\W^\pm_{ij}}$ & $\s,\tS$ & charged baryon $\W^{\N/4}$ or dibaryon, $d_{\SU}\A\A\A$ \\
\cline{2-8}
&\multirow{2}{*}{\parbox{1.5cm}{\centering antisymm\\[-1.5mm] traceless}}  & 
\multirow{2}{*}{$\C_{\Sp}$} & $\Sp(\N_1)\otimes\Sp(\N_2)$ & $\mathbb{Z}_2,\C_{\Sp}$ &$\cgre{\W_{i_1 i_2}}$ & $\s,\tS_1,\tS_2$ & DM exists only in special cases  \\
\cline{4-8}
&&& $\Sp(\N-2)\otimes\Sp(2)$ & $\mathbb{Z}_2,\C_{\Sp}$ &$\cgre{\W}$ & $\s,\tS$ & DM exists only in special cases  \\
\bottomrule
\end{tabular}
\end{small}
\end{table}

\end{landscape}

\begin{table}[t]
	\begin{center}\footnotesize
		\begin{tabular}{l|cc|cc|c}
			\rowcolor[cmyk]{0,0,0.2,0} \multicolumn{6}{c}{Stable DM candidates in the confined phase}\\ \hline
			\rowcolor[cmyk]{0.1,0,0.1,0} Representation & $\SU(\N)$ even & $\SU(\N)$ odd & $\SO(\N)$ even &  $\SO(\N)$ odd & $\Sp(\N)$ even\\
			\hline \hline
			fundamental & $\epsilon \S^\N$, $d\G\G\G$ & $\epsilon \S^\N$, $d\G\G\G$ &$\epsilon \G^{\N/2}$ & $\epsilon \S\G^{(\N-1)/2}$& $\S^T \gamma \S $  \\ \hline
			symmetric & $\epsilon (\G\S)^{\N/2}$, $\epsilon \epsilon \S^\N$, $d\G\G\G$ & $\epsilon \epsilon \S^\N$, $d\G\G\G$ & $\epsilon \G^{\N/2}$ & --- & $\Tr(\S\G)$ \\ \hline
			anti-symmetric & $\epsilon \S^{\N/2}$, $\epsilon \epsilon \S^\N$, $d\G\G\G$ & $\epsilon \epsilon \S^\N$, $d\G\G\G$ & $\epsilon \G^{\N/2}$ & --- & --- \\ \hline
			adjoint & \multicolumn{2}{c|}{$d\G\G\G$, $d\G\S\S$, $f\S\G\G$, $f\S\S\S$} &\multicolumn{2}{c|}{\hbox{see anti-symmetric}} & see symmetric
			\\
			\hline
		\end{tabular}
		\caption{\em\label{tab:stableconf} Stable DM candidates in the confined phase. 
			The --- denotes no accidental DM candidate, 
			either because the action contains $\S^3$ cubics ($\SU$ and $\Sp$ theories) or because di-baryons are unstable ($\SO$ theories). }
	\end{center}
\end{table}

Choosing a given $\G$ and $\S$,
each theory is very predictive, and can have different phases with different
vacuum expectation values $\med{\S}$ and/or condensates, that can break or preserve
its accidental symmetries.
In~\cite{Buttazzo:2019iwr} we  considered scalars in fundamental representations of
SU, SO, Sp and $G_2$ groups, so that there is a unique Higgs phase that shows a
surprising duality with the confined phase.
We here considered scalars in two-index representations of SU, SO, Sp groups,
so that there are two Higgs phases, and qualitatively new patterns of symmetry breaking.
In principle, there could be multiple confined phases,
where different operators acquire condensates.
We found that no duality holds in general, as in some models
the possible breaking patterns of the accidental symmetry
differ  from the Higgs and the condensed phases (giving rise to different or no DM candidates).
Based on the physics behind a Vafa-Witten theorem that applies to gauge interactions of fermions,
we assumed scalar  condensations
that minimally break the symmetries of the theory.\footnote{Interesting physics can arise
in the opposite case.  For example some condensates could break at exponentially small scales
the dark U(1) (giving masses to dark photons) or the Lorentz symmetry in the dark sector.}

For each choice of $\G$ and $\S$ we wrote the most generic renormalizable action
and found its symmetries, as summarized in table~\ref{tab:summary}.
We then wrote its RGEs finding that both Higgs phases can always be realised,
not only with a generic renormalizable potential, but also dynamically
in the specific case of Coleman-Weinberg potentials.
We computed the spectrum in the Higgs phase, first at perturbative level
and next taking into account condensations of the unbroken non-abelian sub-groups.
Each model often leads to DM candidates with a specific phenomenology.
Table~\ref{tab:summary} summarizes our findings in the Higgs phases,
and table~\ref{tab:stableconf} our findings in the confined phases.

Various models give composite DM candidates.
In some models DM is twice composite, as the gauge group $\G$ gets broken to
two non-abelian factors that confine at lower energies.
In most models DM constituents are heavy vectors $\W$ that arise when $\G$ breaks to $\H$.
At the same time, the scalar $\S$ splits into components: we find that they almost never play a role as DM candidates.
In some models DM is made only of the dark gluons of $\H$.

Direct detection rates are often similar to those in~\cite{Buttazzo:2019iwr}, with a main exception.
The symmetry breaking patterns that connect SU with SO or with Sp
($\SU(\N) \to \SO(\N)$ is obtained with $\S$ in the symmetric, 
and $\SU(\N) \to \Sp(\N)$ with $\S$ in the antisymmetric)
feature a spontaneously broken accidental global U(1),
leading to pseudo-Goldstone DM.\footnote{Adding extra fermions charged under both $\G$ and QCD, 
this setup can lead to a QCD axion.
The accidental U(1) is protected by gauge dynamics, 
that  allows only for the U(1) breaking operator $\det\S$ of dimension $\N$.
For large enough $\N$, these axions models
avoid the problem of too large explicit breaking of the U(1) global symmetry,
along the lines of~\cite{1704.01122}.}

Multiple DM candidates arise in theories with multiple accidental symmetries and/or with special relations among 
particle masses that imply extra co-stable states.
As a consequence, the cosmological history that determines the relic DM abundance
is often more complicated than thermal decoupling, involving one
or two re-couplings when broken groups confine.

\footnotesize

\subsubsection*{Acknowledgements}
This work was supported by the ERC grant NEO-NAT.
This work is supported in part by the MIUR under contracts 2017FMJFMW and 2017L5W2PT (PRIN2017). The work of DB is supported in part by the INFN grant FLAVOR.
The work of LDL is supported by the Marie Sk\l{}odowska-Curie Individual Fellowship 
grant AXIONRUSH (GA 840791). 
The work of JWW is supported by
the China Scholarship Council with Grant No. 201904910660.
We thank Claudio Bonati and Franco Strocchi for discussions.

\end{document}